\documentclass[twocolumn,secnumarabic,amssymb, nobibnotes, aps, prl, superscriptaddress, nobalancelastpage,longbibliography]{revtex4-1}

\usepackage{subfigure,dcolumn}
\usepackage[T2A,T1]{fontenc}

\usepackage{epstopdf}
\usepackage{mhchem}
\usepackage{upgreek}

\usepackage{bm}
\usepackage{amsmath} \usepackage{braket} \usepackage{epsfig}
\usepackage{tensor}
\usepackage{CJKutf8}
\usepackage{titlesec}
\usepackage{ifthen}
\usepackage{sidecap}
\usepackage[para,online,flushleft]{threeparttablex}

\usepackage{xr-hyper}
\usepackage{hyperref}
\hypersetup{breaklinks=true,colorlinks=true,linkcolor=blue,citecolor=blue,filecolor=magenta,urlcolor=cyan}

\usepackage{filecontents}
\usepackage[all]{hypcap}

\usepackage{xcolor}
\definecolor{pastelgray}{rgb}{0.81, 0.81, 0.77}
\definecolor{beaublue}{rgb}{0.9, 0.9, 0.93}

\usepackage{listings}
 \usepackage{graphicx}

\renewcommand{\vec}[1]{\mbox{\boldmath $#1$}}

\usepackage{tikz,xcolor,hyperref}

\definecolor{lime}{HTML}{A6CE39}
\DeclareRobustCommand{\orcidicon}{
	\begin{tikzpicture}
	\draw[lime, fill=lime] (0,0) 
	circle [radius=0.16] 
	node[white] {{\fontfamily{qag}\selectfont \tiny ID}};
	\draw[white, fill=white] (-0.0625,0.095) 
	circle [radius=0.007];
	\end{tikzpicture}
	\hspace{-2mm}
}
\foreach \x in {A, ..., Z}{%
	\expandafter\xdef\csname orcid\x\endcsname{\noexpand\href{https://orcid.org/\csname orcidauthor\x\endcsname}{\noexpand\orcidicon}}
}

\begin{document}

\title{Recent Progress in Two-proton Radioactivity}
\author{Long Zhou}
\affiliation{Key Laboratory of Nuclear Physics and Ion-beam Application (MOE), Institute of Modern Physics, Fudan University, Shanghai 200433, China}
\affiliation{Shanghai Research Center for Theoretical Nuclear Physics,
NSFC and Fudan University, Shanghai 200438, China}

\author{Si-Min Wang \orcidB{}}\email{wangsimin@fudan.edu.cn}
\affiliation{Key Laboratory of Nuclear Physics and Ion-beam Application (MOE), Institute of Modern Physics, Fudan University, Shanghai 200433, China}
\affiliation{Shanghai Research Center for Theoretical Nuclear Physics,
NSFC and Fudan University, Shanghai 200438, China}

\author{De-Qing Fang \orcidC{}}\email{dqfang@fudan.edu.cn}
\affiliation{Key Laboratory of Nuclear Physics and Ion-beam Application (MOE), Institute of Modern Physics, Fudan University, Shanghai 200433, China}
\affiliation{Shanghai Research Center for Theoretical Nuclear Physics,
NSFC and Fudan University, Shanghai 200438, China}

\author{Yu-Gang Ma \orcidD{}}\email{mayugang@fudan.edu.cn}
\affiliation{Key Laboratory of Nuclear Physics and Ion-beam Application (MOE), Institute of Modern Physics, Fudan University, Shanghai 200433, China}
\affiliation{Shanghai Research Center for Theoretical Nuclear Physics,
NSFC and Fudan University, Shanghai 200438, China}

\begin{abstract}
During the last few decades, rare isotope beam facilities have provided unique data for studying the properties of nuclides located far from the beta-stability line. Such nuclei are often accompanied by exotic structures and radioactive modes, which represent the forefront of nuclear research. Among them, two-proton (2$p$) radioactivity is a rare decay mode found in a few highly proton-rich isotopes. The $2p$ decay lifetimes and properties of emitted protons hold invaluable  information regarding the nuclear structures in the presence of a low-lying proton continuum; as such, they have attracted considerable research attention. In this review, we present some of the recent experimental and theoretical progress regarding the $2p$ decay, including technical innovations for measuring nucleon--nucleon correlations and developments in the models that connect their structural aspects with their decay properties. This impressive progress should play a significant role in elucidating the mechanism of these exotic decays, probing the corresponding components inside nuclei, and providing deep insights into the open quantum nature of dripline systems.
\end{abstract}

\keywords{Exotic decay, Two-proton radioactivity, Nucleon--nucleon correlation, Experimental and theoretical development}

\maketitle

\section{Introduction}

A unique laboratory for many-body quantum physics, nuclei are composed of positively-charged protons and neutral neutrons; they offer invaluable data for testing fundamental theories and studying the universal properties of fermionic systems \cite{Cooper1959,Leggett2004,Brink2005,Broglia2013,Dean2003}. Of around the 7000 atomic nuclei thought to exist \cite{Erler2012, Neufcourt2020}, less than half have been experimentally observed; these include the 286 primordial nuclides located in the $\beta$-stability valley. Meanwhile, the rare isotopes that inhabit remote regions of the nuclear landscape (around and beyond the particle driplines) are often accompanied by exotic structures and decay modes, owing to the weak binding and strong continuum coupling. Such nuclei are also referred to as open quantum systems, and they represent the forefront of nuclear structure and reaction research \cite{Dobaczewski1998,RISAC,Dobaczewski2007,Forssen2013,Balantekin2014,NSACLRP2015}.

\begin{figure*}[htb]
\includegraphics[width=0.9\textwidth]{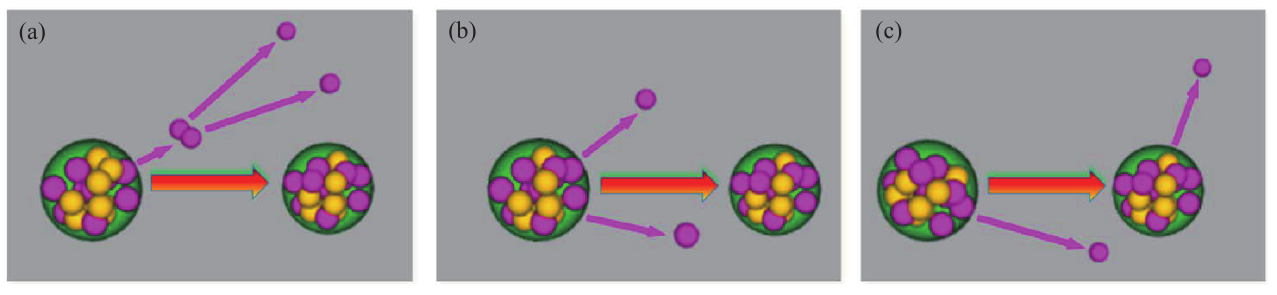}
\caption[T]{
Sketch of the different $2p$ emission mechanisms: (a) diproton emission, (b) three-body (large-angle) emission, and (c) sequential emission \cite{Fang2020}.
\label{Fang2020}}
\end{figure*}

Among the most fundamental properties of such exotic nuclear systems, the first to be established are---in general---the possible radioactive modes and corresponding lifetimes. When approaching the edge of nuclear stability, new decay modes arise, including the exotic two-proton (2$p$) radioactivity. The concept of 2$p$ decay is relatively old; it was proposed by Goldansky in 1960 \cite{Goldansky1960}. In such decay cases, the single-proton emission is energetically forbidden or strongly suppressed by the inner structure and angular momentum selection rule \cite{Goldansky1960,Pfutzner2004,Blank2008,Pfutzner2012,Pfutzner2013,Giovinazzo2013,Olsen2013,*Olsen2013Sep}. Experimental studies into $2p$-decay began in the 1970s and initially focused upon light nuclei such as $^6$Be \cite{Geesaman1977,Bochkarev1989} or $^{12}$O \cite{KeKelis1978,Kryger1995}. However, owing to their short lifetimes and broad intermediate states, the decay mechanisms of these nuclei are not fully understood. The first direct observation of 2$p$ emission from a long-lived ground state was achieved for $^{45}$Fe via projectile fragmentation and in-flight identification~\cite{Giovinazzo2002,Pfutzner2002}. So far, researchers have identified only a very small number of nuclei that can decay by emitting two protons from their ground and/or excited states. 

The $2p$ decay lifetimes and properties of emitted protons carry invaluable information for nuclear physics research. In contrast to $\beta$ and $\gamma$ decays (which are mainly governed by weak and electromagnetic interactions, respectively), $2p$ decays arise from the competition between strong nuclear forces and electromagnetic interactions. As a result, the corresponding lifetimes span a wide range of time scales; this offers a valuable testing ground for exploring unified theories and elucidating the interplay between these two types of fundamental interactions.

Moreover, the $2p$ emission process, which typically involves one core and two protons, is a three-body process. In the presence of a low-lying continuum, this feature produces difficulties and increases the investigation complexity compared to the one-proton emission. In general, the $2p$ decay mechanisms can be roughly divided into three types (see Fig.\,\ref{Fang2020} and Ref.\,\cite{Blank2008}): (i) Diproton emission, an extreme case in which two strongly correlated protons are emitted in almost identical directions; here, the diproton emission is essentially two protons constrained by pair correlation in a quasi-bound $s$-singlet (i.e., $^{1}\mathrm{S}_{0}$) configuration; because of the Coulomb barrier, such quasi-bound states can only exist for a short while before becoming separated after penetrating through the barrier. (ii) Three-body (large-angle) simultaneous emission, in which the core is separated from two protons simultaneously, and the valence protons are positioned remotely in coordinate space and emitted with large opening angles. (iii) Two-body sequential emission, in which the mother nucleus first emits a proton to 
an intermediate state of the neighboring nucleus, which itself then emits another proton to the final state, as shown in Fig.~\ref{Fang2020}(c). The sequential decay can be treated as a two-step $1p$ decay; hence, the first two emission mechanisms typically attract more interest, owing to their three-body character
 in the presence of low-lying continuum. 
In this three-body process, the relative momenta and opening angles between the two emitted protons contain the specific form of the nucleon wave function as well as the interactions between nucleons; hence, such processes are useful for studying the structures, decay properties, and nucleon--nucleon pairing correlations (in particular, the $pp$ correlation) of nuclei around the proton dripline \cite{Catara1984,Pillet2007,Hagino2005,Hagino2007,Hagino2014,Matsuo2012,Fossez2017}. In addition, these processes offer a good method for investigating the astro-nuclear $(2p,\gamma)$ and $(\gamma,2p)$ processes, which are closely related to the waiting point nuclei \cite{Fisker2004}. Meanwhile, the structures and reactions of certain $2p$ emitting nuclei are very important for understanding the cycle processes of element synthesis (e.g., NeNa and NaMg cycles) in nuclear astro-reactions \cite{Seweryniak2005}. 

Therefore, such processes pose a unique opportunity and daunting challenge for both experimental and theoretical studies. Experimentally, the utility of implantation into an Si array increases the efficiency of $2p$ emission observations \cite{Blank2008,Pfutzner2012}, and the development of time-projection chambers \cite{Giovinazzo2007} has allowed researchers to trace the decay properties and asymptotic correlations of the emitted valence protons. In addition, from these properties and information regarding the intermediate state, the structure and decay mechanisms of the mother nucleus can be investigated. Meanwhile, these high-quality $2p$ decay data obtained from exotic beam facilities also require the development of comprehensive theoretical approaches capable of simultaneously describing the structural and reaction aspects of this exotic decay problem~\cite{Blank2008,Pfutzner2012}.

Because only a few 2$p$ emitters have been identified, the mechanisms and properties of this exotic decay are largely unknown. To search for 2$p$ emitter candidates, systematical calculations have been performed using various theoretical models ~\cite{Nazarewicz1996,Ormand1997,Barker2001,Brown2003,Goncalves2017,Olsen2013,*Olsen2013Sep,Cui2020,*Cui2021,Neufcourt2020_2,LiuHM2021,Delion2022}. In these models, most predictions are made by treating the internal and asymptotic regions of the nucleus separately; this is efficient for estimating certain bulk proprieties (e.g., binding energy) and for spectroscopy. However, these 2$p$ emitters are located across vast regions of the nuclear landscape; thus, they have very different structures and decay properties. Meanwhile, this complicated three-body process is a consequence of the interplay between the internal nuclear structure, asymptotic behavior, and continuum effect. Therefore, to give a comprehensive description of a decay process (in particular, for the mechanism and asymptotic nucleon--nucleon correlations originating from the internal nuclear structure and distorted by the long-range Coulomb interaction), fine-tuned studies are required for each individual system. This becomes especially challenging for 2$p$ decays because the Coulomb barrier strongly suppresses the wave function at large distances; this also makes the 2$p$ lifetime relatively sensitive to the low-$\ell$ wave function components inside the nucleus.

Over the past few decades, impressive progress has been made (in both experimental and theoretical studies) in these directions; much of this has been well reviewed in Refs.\,\cite{Blank2008,Pfutzner2012}. In this review, we present some of the recent research, including technical innovations for measuring nucleon--nucleon correlations and model developments for connecting structural aspects and decay properties. Meanwhile, numerous open questions (e.g., the $2p$ decay mechanism and impacts of internal structure) remain under debate. Further studies may help provide deep insights into the $2p$ decay process as well as other open quantum systems.

\section{Brief History}

\begin{figure*}[!htb]
\includegraphics[width=0.8\textwidth]{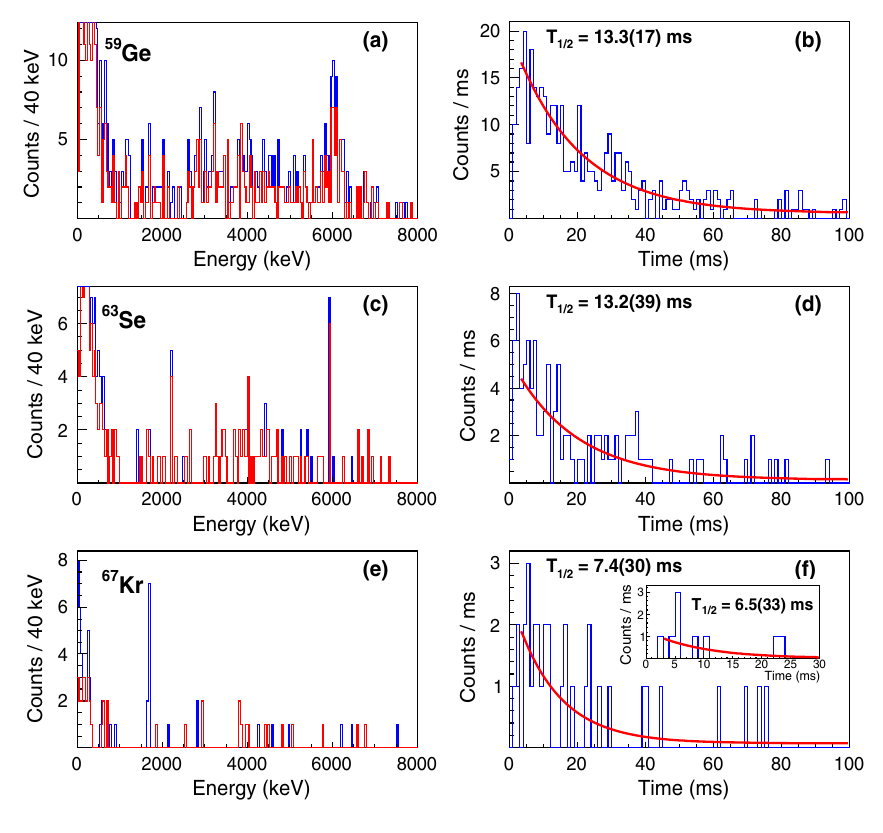}
\caption[T]{
The energy spectra (left) and decay-time distributions (right) of $^{59}\mathrm{Ge}$ [(a), (b)], $^{63}\mathrm{Se}$ [(c), (d)], and $^{67}\mathrm{Kr}$ [(e), (f)]. For the charged-particle spectra, the blue histograms correspond to all decay events, and the red ones denote those events which coincid with the detection of $\beta$-decay particles in neighboring detectors. The 1690\,keV peak of $^{67}\mathrm{Kr}$ is attributable to 2$p$ decay. The inset in panel (f) shows the half-life of $^{67}\mathrm{Kr}$, as determined from the events in the 1690\,keV peak. See Ref.\,\cite{Goigoux2016} for details.
\label{Goigoux2016}}
\end{figure*}

\begin{figure}[!htb]
\includegraphics[width=0.9\linewidth]{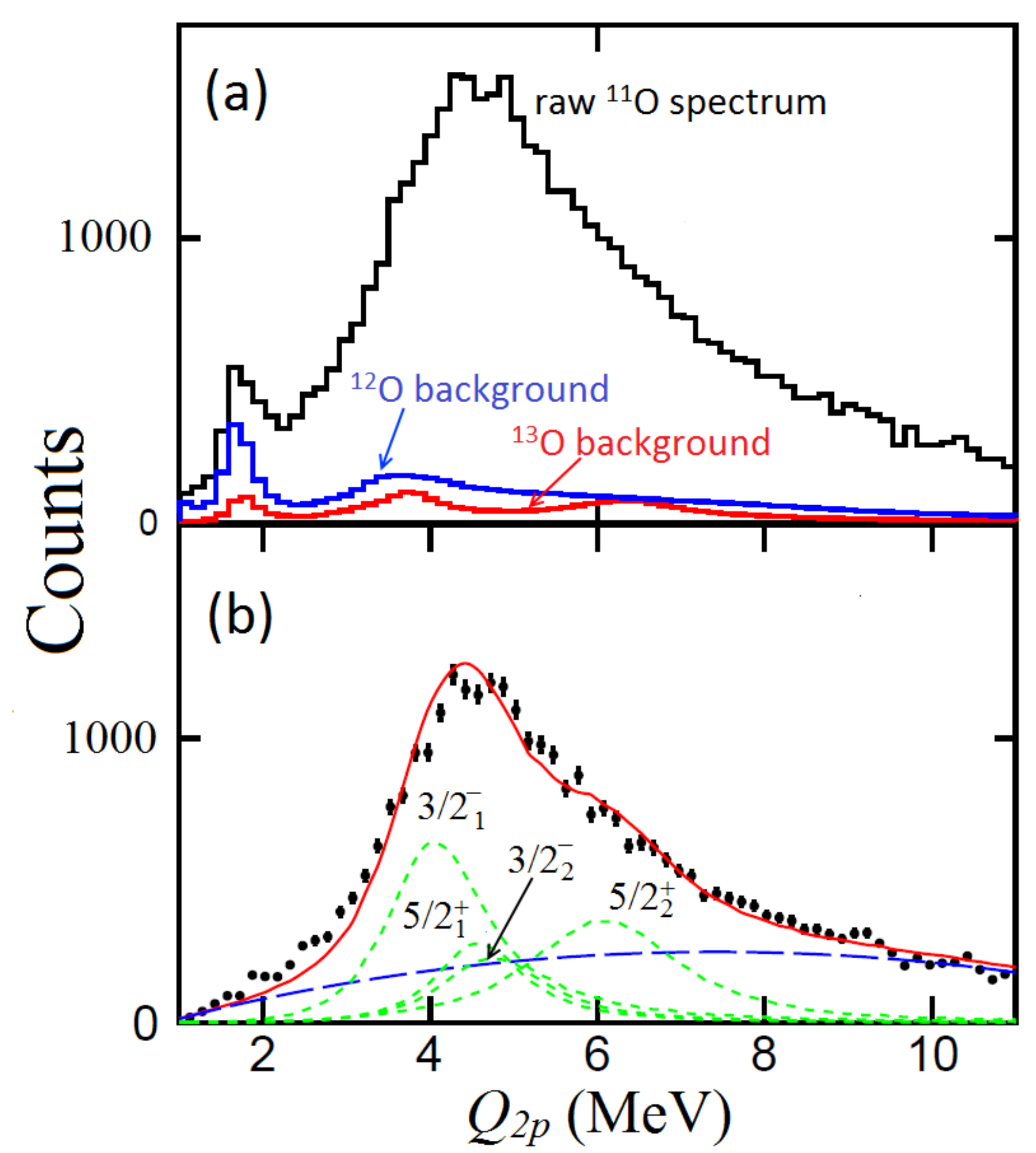}
\caption{Spectrum for the 2$p$ decay energy $Q_{2p}$ of $^{11}$O,  as reconstructed from detected 2$p$+$^{9}$C events (a) including contamination from $^{10, 11}$C events and (b) with the contamination removed. The solid curve in (b) is a fit to the data containing contributions from four low-lying states (short-dashed curves), as predicted by Gamow coupled-channel calculations \cite{Webb2019,Wang2019}; the long-dashed curve is the fitted background. See Ref.\,\cite{Webb2019} for details.}
\label{Webb2019}
\end{figure}

In the early 1960s, Goldansky suggested a novel radioactive mode of $1p$ (one-proton) or $2p$ emission in proton-rich nuclei far from the stability line \cite{Goldansky1960,Goldansky1961,Blank2008}. The earliest experimental discovery of such phenomena was the single-proton emission of excited state nuclei. In 1963, a $\beta$-delayed proton emission was believed to have been identified from $^{25}\mathrm{Si}$; however, this turned out to originate from the excited state of $^{25}\mathrm{Al}$ \cite{Barton1963}. Later, the proton emission was discovered in the isomer state of $^{53}\mathrm{Co}$ \cite{Jackson1970} and the ground state of $^{151}\mathrm{Lu}$ \cite{Hofmann1982}, in 1970 and 1982, respectively. Since then, more than 20 nuclides have been confirmed as the ground-state proton emitters; this offers invaluable information with which to study the nuclear structure.

{\it $2p$ decay from excited states.} --- The ground-state $2p$ emission occurs in a region even farther beyond the proton dripline. The candidates given in the early theories mainly included $^{39}\mathrm{Ti}$, $^{42}\mathrm{Cr}$, $^{45}\mathrm{Fe}$, and $^{48,49}\mathrm{Ni}$ \cite{Brown1991,*Brown1991_2,Cole1996,Ormand1996}. It is very difficult to produce these nuclei in experiments; hence, this exotic phenomenon has not been found, despite years of research. In 1980, Goldansky postulated a $\beta$-delayed $2p$ emission from excited nuclei 
and proposed that the candidates could be proton-rich nuclei with charge numbers $Z$ ranging from 10 to 20 \cite{Goldansky1980}. The first observation of $\beta$-delayed $2p$ emission from $^{22}\mathrm{Al}$ was achieved at Lawrence Berkeley National Laboratory (LBNL) in 1983 \cite{Cable1983}; here, the transition proceeded via the $T=2$ isobaric-analog state ($E_x$ = 14.044\,MeV) in  $^{22}\mathrm{Mg}$. Inspired by this experiment, further $\beta$-delayed $2p$ emission nuclei were investigated, including $^{23}\mathrm{Si}$, $^{26}\mathrm{P}$, $^{27}\mathrm{S}$, $^{31}\mathrm{Ar}$, $^{35}\mathrm{Ca}$, $^{39}\mathrm{Ti}$, $^{43}\mathrm{Cr}$, $^{46}\mathrm{Fe}$, and $^{50,51}\mathrm{Ni}$ \cite{Blank1997,Honkanen1983,Borrel1991,Reiff1989,Aysto1985,Moltz1992,Borrel1992,Dossat2007,Audirac2012,Pomorski2014}. 
The excited $2p$ emitters might not only be produced by $\beta$ decay but also by nuclear reactions such as $^{14}\mathrm{O}$ or  $^{17,18}\mathrm{Ne}$ \cite{Bain1996,Chromik1997,delCampo2001,Raciti2008} and may include fragmentation, transfer, and pick-up reactions. Owing to the large decay energy, the lifetime of a $2p$ emission from the excited state is generally short, about the order of ${10}^{-21}$\,s, much shorter than those from the ground states. Consequently, the corresponding decay mechanism is harder to identify in these excited $2p$ emission processes. In recent years, radioactive beam devices have been used to conduct a series of experimental measurements, including the $\beta$-delayed $2p$ emission of $^{22}\mathrm{Al}$,$^{22,23}\mathrm{Si}$, and $^{27}\mathrm{S}$, and the $2p$ emission of excited $^{22}\mathrm{Mg}$,$^{23}\mathrm{Al}$,$^{27,28}\mathrm{P}$, and $^{28,29}\mathrm{S}$ nuclei \cite{WangYT2018,Xu2017,WangK2018,Ma2015,Fang2016,Xu2013,Lin2009,Xu2010,Shi2021,Wu2021}; this constitutes important progress in the characterization of excited $2p$ emitters.

{\it $2p$ decay from ground states.} --- The ground-state $2p$ emitters are typically located beyond the proton dripline, and they are extremely difficult to produce. Therefore, researchers initially investigated the light-mass region of the nuclear landscape \cite{Geesaman1977,Kryger1995,Mukha2010,Mukha2007,Webb2019}. Owing to a large proton--neutron asymmetry, these short-lived $2p$ emitters (e.g., $^{6}\mathrm{Be}$,$^{11,12}\mathrm{O}$,$^{16}\mathrm{Ne}$, and $^{19}\mathrm{Mg}$) tend to have short lifetimes and broad intermediate states, resulting in more complicated decay dynamics (see the discussion below for details). These very unstable systems are usually accompanied by exotic structures, which makes them harder to study experimentally and theoretically. For instance, as shown in Fig.\,\ref{Webb2019}, the mirror of the halo nucleus $^{11}$Li, $^{11}\mathrm{O}$, was recently found to exhibit $2p$ decay with a ground-state decay width of more than 1\,MeV \cite{Webb2019,Webb2020}. Many efforts---both theoretical and experimental---have been made to investigate their decay mechanisms and the impact of their internal nuclear structures; this has resulted in the excellent progress detailed below.

Meanwhile, unlike the general situation for light-mass $2p$ emitters, the criteria originally introduced by Goldansky \cite{Goldansky1960,Goldansky1988} for the $2p$ decay was $Q_{2p}$ > 0 and $Q_{p}$ < 0, where $Q$ denotes the decay energy. In this case, the one-proton decay channel is energetically forbidden. The nuclei satisfying these requirements are typically in the mid-mass region, in which both mother and neighboring nuclei can be quasi-bound. Until 2002, the $2p$ emission from the long-lived ground state was directly observed for the first time in $^{45}\mathrm{Fe}$ \cite{Giovinazzo2002,Pfutzner2002}. The experiment was conducted at the Grand Accelerateur National d’Ions Lourds (GANIL) in France and the Gesellschaft f\"ur Schwerionenforschung (GSI) in Germany, in which $12$ and $3$ $2p$ decay events were observed, respectively. The extracted half-lives were $4.7_{-1.4}^{+3.4}$ and $3.2_{-1.0}^{+2.6}$\,ms, respectively. Later, in 2005, a $2p$ emission from the ground state of $^{54}\mathrm{Zn}$ was identified in an experiment conducted by GANIL \cite{Blank2005}, which observed seven events and measured a half-life of $3.7_{-1.0}^{+2.2}$\,ms. In the same year, GANIL studied the decay of the ground state of $^{48}\mathrm{Ni}$ \cite{Dossat2005}: one of the four decay events was entirely compatible with $2p$ radioactivity. This was later confirmed by Michigan State University (MSU). The corresponding experiment was performed at the National Superconducting Cyclotron Laboratory (NSCL/MSU), with four $2p$ events being directly observed and a half-life of $2.1_{-0.4}^{+1.4}$\,ms obtained \cite{Pomorski2011}. Further  experiments have been carried out or proposed, to more deeply probe the properties of this exotic decay mode in $^{45}\mathrm{Fe}$, $^{48}\mathrm{Ni}$, and $^{54}\mathrm{Zn}$ \cite{Miernik2009,Pomorski2014,Ascher2011}. In 2016, $^{67}\mathrm{Kr}$ was measured with a half-life of $7.4(30)$\,ms in an experiment performed at the Institute of Physical and Chemical Research (RIKEN) (see Fig.
\,\ref{Goigoux2016} and Ref.\,\cite{Goigoux2016}). By considering the dead-time correction of the acquisition system, ~13 $2p$ events were detected, in which the branching ratio was 37(14)\%. This makes $^{67}\mathrm{Kr}$ the heaviest $2p$ emitter ever discovered. So far, only these four long-lived $2p$ emitting ground-state nuclei have been identified. Other candidates are theoretically possible, including $^{59}\mathrm{Ge}$ and $^{63}\mathrm{Se}$, which are predicted to be $2p$ emitters. However, in the experimental measurements \cite{Goigoux2016},
$^{59}\mathrm{Ge}$ and $^{63}\mathrm{Se}$ are mainly observed via $\beta$-delayed proton(s) emission, and no $2p$ decay phenomenon have been observed.

\begin{table}[!htb]
\caption{The currently confirmed ground-state 2$p$ emitters. Also shown are the corresponding 2$p$ decay energies (in MeV) and decay widths (half-lives). Experimental values are from Refs.\,\cite{ENSDF,Webb2020,Wamers2014,Jin2021,Pomorski2014,Goigoux2016,Ascher2011}}
\begin{ruledtabular}
\begin{tabular}{ l  c  c  }
\text { Nuclei }  & $Q_{2 p}$ & $\Gamma$ / T$_{1/2}$ \\
\hline
${ }_{4}^{6} \mathrm{Be}$ \cite{Geesaman1977,Bochkarev1989,Egorova2012} &  1.372(5) & 92(6)\,keV \\
${ }_{6}^{8} \mathrm{C}$ \cite{Charity2010} &  2.111(19) & 230(50)\,keV \\
${ }_{~8}^{11} \mathrm{O}$ \cite{Webb2019} &  4.25(6) & 2.31(14)\,MeV \\
${ }_{~8}^{12} \mathrm{O}$ \cite{KeKelis1978,Kryger1995,Webb2019_2} &  1.737(12) & 0.40(25)\,MeV \\
${ }_{10}^{15} \mathrm{Ne}$ \cite{Wamers2014} & 2.522(66) & 0.59(23)\,MeV \\
${ }_{10}^{16} \mathrm{Ne}$ \cite{Mukha2008,Mukha2010,Brown2014,Brown2015,KeKelis1978} & 1.401(20) & $\le$ 80\,keV \\
${ }_{12}^{18} \mathrm{Mg}$ \cite{Jin2021} & 3.44 & 115(100)\,keV \\
${ }_{12}^{19} \mathrm{Mg}$ \cite{Mukha2007,Voss2014,Brown2017,Xu2018} & 0.76(6) & 4.0(15)\,ps \\
${ }_{18}^{30} \mathrm{Ar}$ \cite{Mukha2015,Xu2018} & 3.42(8) & $\le$ 10\,ps \\
${ }_{26}^{45} \mathrm{Fe}$ \cite{Giovinazzo2002,Pfutzner2002} & 1.80(20) & 2.45(23)\,ms \\
${ }_{28}^{48} \mathrm{Ni}$ \cite{Dossat2005,Dossat2007,Pomorski2011,Pomorski2014} & 1.26(12) & 2.1$^{+1.4}_{-0.4}$\,ms \\
${ }_{30}^{54} \mathrm{Zn}$ \cite{Blank2005,Ascher2011} & 1.28(21) & 1.59$^{+0.60}_{-0.35}$\,ms \\
${ }_{36}^{67} \mathrm{Kr}$ \cite{Goigoux2016} & 1.690(17) & 7.4(30)\,ms \\
\end{tabular}
\end{ruledtabular}
\end{table}

\section{Experimental Progress}

The $2p$-emitting nuclei are likely to inhabit remote regions of
the nuclear landscape, regardless of their energy state; thus, the production rate is always very low, which makes the experimental conditions difficult to realize. Consequently, only a few $2p$ emitters have been identified, through rare events. Although the decay process can also be determined via other physical quantities (e.g., decay energy and half-life) to clearly identify the different mechanisms of $2p$ emission, it is necessary to measure sufficient $2p$ emission events, as well as to measure the correlation (including the relative momentum and opening angle) between the emitted core and two protons. To this end, many laboratories around the world are developing next-generation facilities and state-of-the-art detectors, to explore candidate $2p$ emitters and their properties. In this section, we briefly introduce some of the recent experimental progress.

\subsection{Measurement in experiments}

Generally, the half-lives of proton-rich nuclei decrease under the increase in proton--neutron asymmetry, and the span ranges from ${10}^{1}$\,s to ${10}^{-22}$\,s. Therefore, different production mechanisms, detection methods, and technologies \cite{WangHL2022,Bai2022,Nan2021,MaCW2021,MaCW2022} are used for such nuclei in experiments, depending on the half-life. In early years, light particle-emission nuclei (e.g., $^{6}\mathrm{Be}$ and $^{12}\mathrm{O}$) were produced by transfer reactions with protons or $^{3}\mathrm{He}$ beams \cite{Geesaman1977,Kryger1995}. Nowadays, the production of exotic nuclei is performed using two main technologies:(i) in-flight projectile fragmentation (PF) and (ii) isotope separation on-line (ISOL). A PF radioactive beam device using heavy ions as the incident beam was installed in Berkeley Laboratory in the 1970s and has been widely used since then; this allows heavier nuclei to be accelerated \cite{Symons1979}. In this method, the short-lived radionuclides are produced via fragmentation of the incident nucleus on a thin target. The momentum direction of the produced radionuclide is strongly consistent with the direction of the incident beam. Then, the generated radionuclides is selected and transferred via a fragment separator to a secondary reaction zone, where different experiments can be performed using different secondary target and detector setups. PF can produce nuclei with very short half-lives or close to the proton dropline; however, the secondary-beam quality is relatively low. When studying the decay of nuclei with very short half-lives, the primary detection method is the in-flight decay one, because most nuclei decay before they can pass through the separation and purification spectrometer. ISOL is a radioactive nuclear beam device developed in the 1950s; it was first proposed and verified by O. Kofoed-Hansen and K.O. Nielsen of Copenhagen University, Denmark, in 1951 \cite{Kofoed1951}. They bombarded a uranium target with a neutron beam, causing it to fragment and produce the neutron-rich nuclides $^{89,90,91}\mathrm{Kr}$ and $^{89,90,91}\mathrm{Rb}$. Then, the generated radionuclides were diffused and extracted to the ion source for ionization, by heating the thick target. Subsequently, the ionized radionuclides were extracted for mass spectrometry and electromagnetic separation processes; finally, the separated and purified radioactive beams were transferred into the secondary reaction zone to perform the experiment. In ISOL, the generation, ionization, separation, and experiment procedures of the radioisotopes are carried out continuously. One attractive feature of ISOL facilities is their extraordinarily good secondary-beam quality, which offers a very low cross-contamination (below ${10}^{-4}$); however, it is difficult to produce short-lived nuclides. The main detection method for the decay of nuclei produced by ISOL is the implantation method, because the half-life is considerably longer than the flight time in the spectrometer after generation. The measurement methodology also depends on different circumstances, including the half-life, decay mechanism, and experimental objective. Here, we use $2p$ emission as an example to demonstrate the advantages and disadvantages of different measurement techniques, as well as the corresponding scopes of applications.

{\it In-flight decay technique.} --- As discussed, in-flight decay constitutes an effective method that can be used to measure the decay of $2p$ emitters with half-lives of less than a pico-second. In this case, the unstable nuclei decay shortly after being produced by the radioactive beam facilities. Among the ground-state $2p$ emitters discovered thus far, $^{19}\mathrm{Mg}$ was investigated using this technique \cite{Mukha2007}. Meanwhile, it also consistutes the main experimental technique for studying excited $2p$ emissions. 

\begin{figure*}[!htb]
\includegraphics[width=0.7\textwidth]{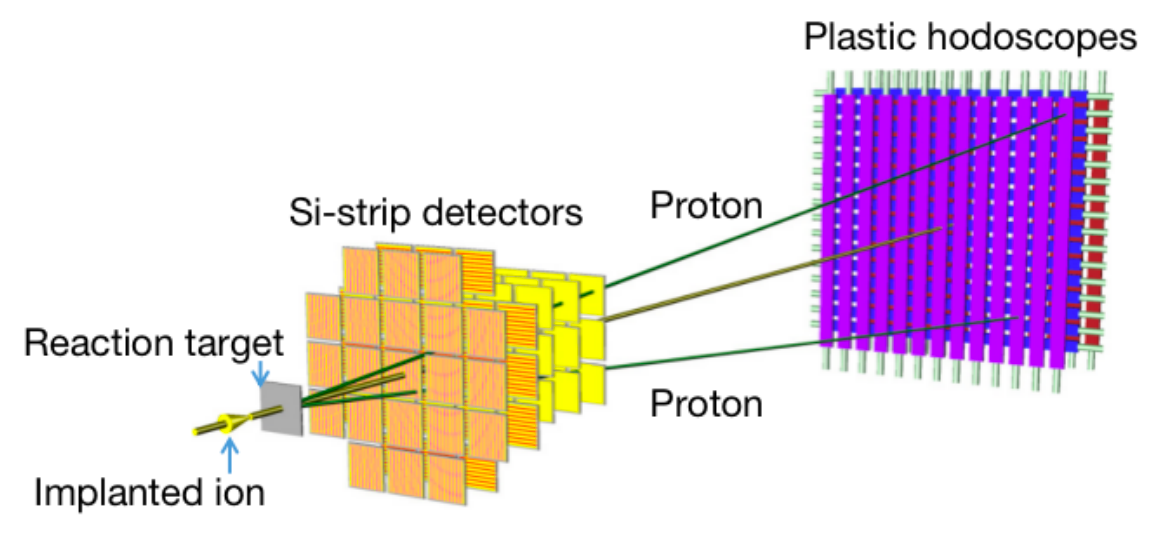}
\caption[T]{
Sketch of the detector setup for in-flight decay. See Refs\,\cite{Ma2015,Fang2016} for details.
\label{Ma2015}}
\end{figure*}

In the following, the $2p$ emission of excited state $^{22}\mathrm{Mg}$ \cite{Ma2015} is introduced as an application of in-flight decay for measuring the decay of very short-lived proton-rich nuclei. The experiment was performed using the projectile fragment separator (RIPS) beam line at the RI Beam Factory (RIBF), operated by the RIKEN Nishina Center and the Center for Nuclear Study, University of Tokyo. A primary beam of 135\,A\,MeV $^{28}\mathrm{Si}$ was used to produce secondary $^{23}\mathrm{Al}$ and $^{22}\mathrm{Mg}$ beams with incident energies of 57.4\,A\,MeV and 53.5\,A\,MeV, respectively, in the center of the carbon reaction target. Beyond the reaction target, five layers of silicon detectors and three layers of plastic hodoscopes were positioned as shown in Fig.~\ref{Ma2015}. The first two layers of Si-strip detectors were located ~50\,cm downstream of the target and used to measure the emission angles of the fragment and protons. Three layers of 3$\times$3 single-electrode Si were used as ${\Delta}{E}$-$E$ detectors for the fragment. The three layers of plastic hodoscopes were located ~3\,m downstream of the target and were used as ${\Delta}{E}$ and $E$ detectors for protons. The time of flight of the protons was measured using the first layer. Clear particle identifications were obtained using this setup for the kinematically complete three-body decays. The momenta and emission angles for protons and the residue were determined by analyzing the detector signals. The excitation energy (${E}^{*}$) of the incident nucleus was reconstructed from the difference between the invariant mass of the three-body system and the mass of the mother nucleus in the ground state; this provides useful information with which to determine the decay mechanism of $2p$ emission. These experimental studies indicate that the highly excited state of $^{22}\mathrm{Mg}$ may produce a diproton emission \cite{Ma2015}.

\begin{figure*}[!htb]
\includegraphics[width=0.7\textwidth]{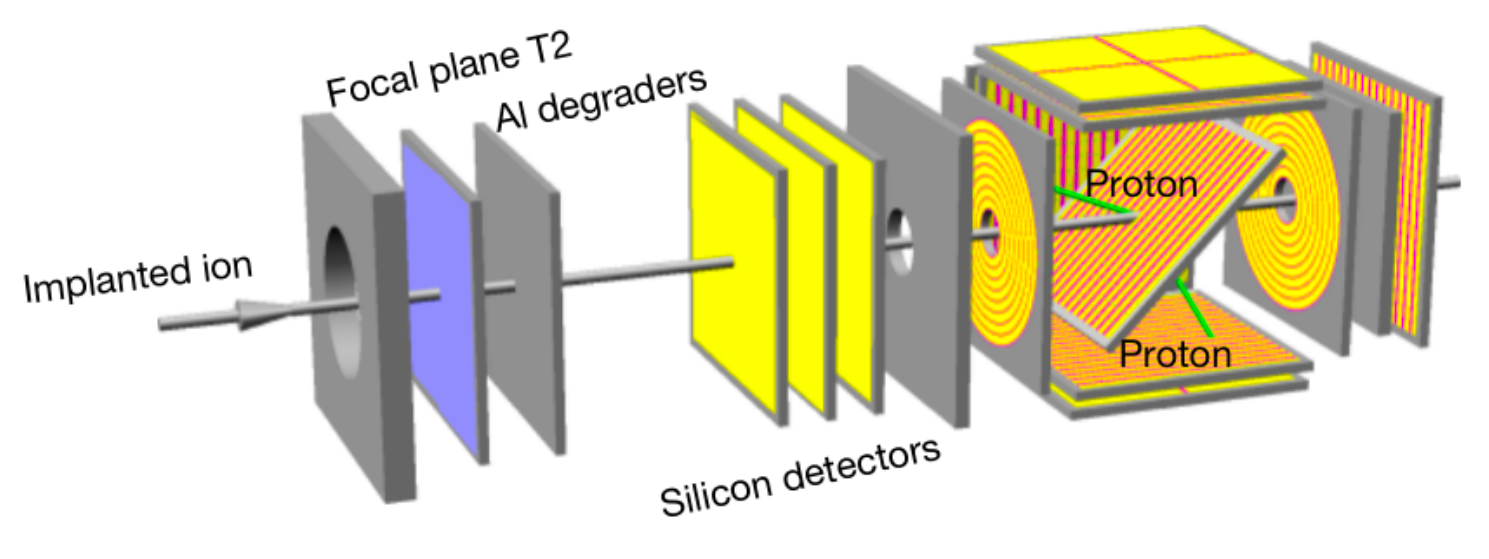}
\caption[T]{
Sketch of the detector setup for implantation decay. See Ref.\,\cite{WangYT2018} for details.
\label{WangYT2018}}
\end{figure*}

{\it Implantation decay method.} --- Implantation decay is one of the most popular methods used to experimentally measure long-lived nuclei. In terms of $2p$ decay, it is particularly useful for the $\beta$-delayed decays of proton-rich nuclei, because of the long half-life of $\beta$ decay (as governed by the weak interaction). As an example, we introduce an experiment involving the $\beta$-delayed $2p$ emission of $^{22}\mathrm{Al}$, performed at the Radioactive Ion Beam line in Lanzhou (RIBLL) \cite{WangYT2018,WangYTNST2018,Sun2015}. A sketch of the detector setup for implantation decay is shown in Fig.~\ref{WangYT2018}. 
The emitted nucleus $^{22}\mathrm{Al}$ produced by RIBLL first passes through two scintillation detectors (located at the first and second focal planes T1 and T2, respectively). The energy of the secondary beam is adjusted by selecting a combination of aluminum degraders with different thicknesses through the stepper motors; thus, most of the $^{22}\mathrm{Al}$ nuclei can be injected and halted at the decay position. Three silicon detectors with thicknesses of 300\,$\mu$m are installed after the aluminum degraders, to measure the energy loss (${\Delta}{E}$) of the beam ions. The identification of incident ions is performed using a ${\Delta}{E}$-TOF. Then, the incident ions are injected and stopped using a double-sided silicon strip detector (DSSD) set with a $42.5^{\circ}$ tilt angle, to increase the effective thickness. The subsequent decays are measured and correlated to the preceding implantations using the position and time information, and the emitted charged particles with sufficient energies to escape the centered DSSD are measured by the detection array, which is composed of four DSSDs and four quadrant silicon detectors (QSDs). In addition, five high-purity Germanium (HPGe) $\gamma$-ray detectors outside the silicon detectors are used to detect the $\gamma$-rays emitted by the decay, and several QSDs are also installed to measure anti-coincidences with $\beta$ particle detections. In the implantation decay experiment, the radioactive nuclei are injected into silicon detectors and then decay via $\beta$-delayed $1p$ or $2p$ emission after a certain time. Both processes produce corresponding signals in the silicon detector. Moreover, it should be noted that the energy loss of the radionuclide in silicon detectors is relatively high (several hundreds of MeV) compared to the energy of protons emitted during the decay (generally only a few MeV). The different energy scales makes it difficult to measure the signals with the same magnification. To solve this problem, each output channel of the preamplifiers for the three DSSDs is split into two parallel electronic chains with low and high gains, to measure both the implantation events with energies up to hundreds of MeV and the decay events with energies on the order of hundreds of keV or less. One uses a relatively small amplification factor to measure the signal of injected ions; the other one measures the emitted protons. Owing to the granularity and position resolution of DSSDs, the opening angle and relative momentum correlations of two emitted protons can be easily reconstructed. Moreover, the decay channel and half-life can also be obtained. The decay half-life corresponds to the entire process, including the $\beta$ decay \cite{WangYT2018}.

\begin{figure}[htb]
\includegraphics[width=1.0\columnwidth]{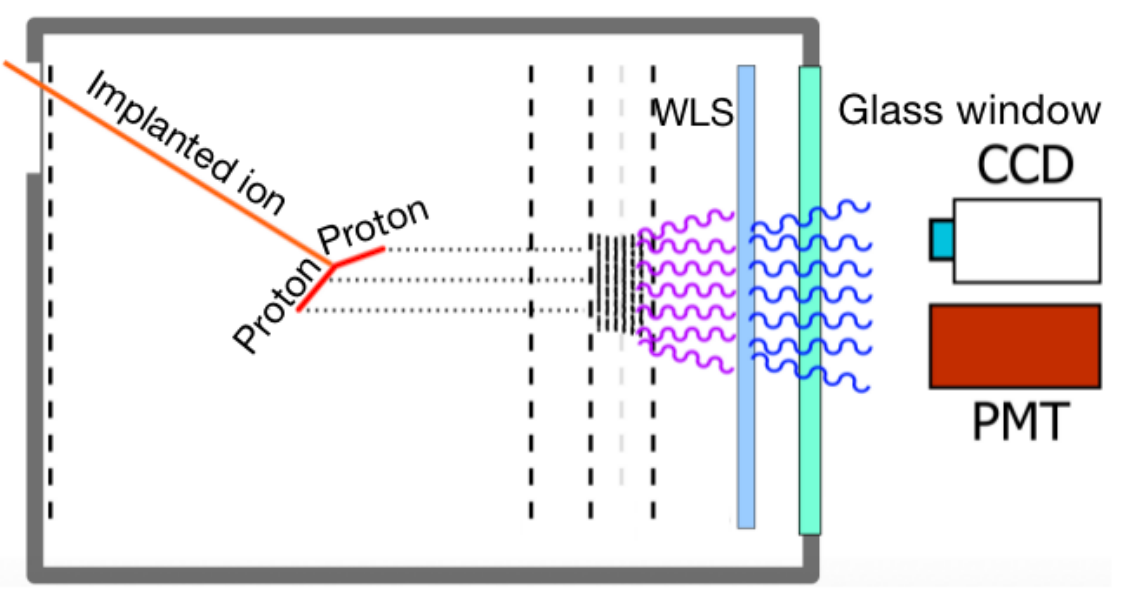}
\caption[T]{
Sketch of the detector setup for the optical time projection chamber-charge coupled device (OTPC-CCD) \cite{Miernik2009}.
\label{Miernik2009}}
\end{figure}

\begin{figure*}[htb]
\includegraphics[width=0.8\textwidth]{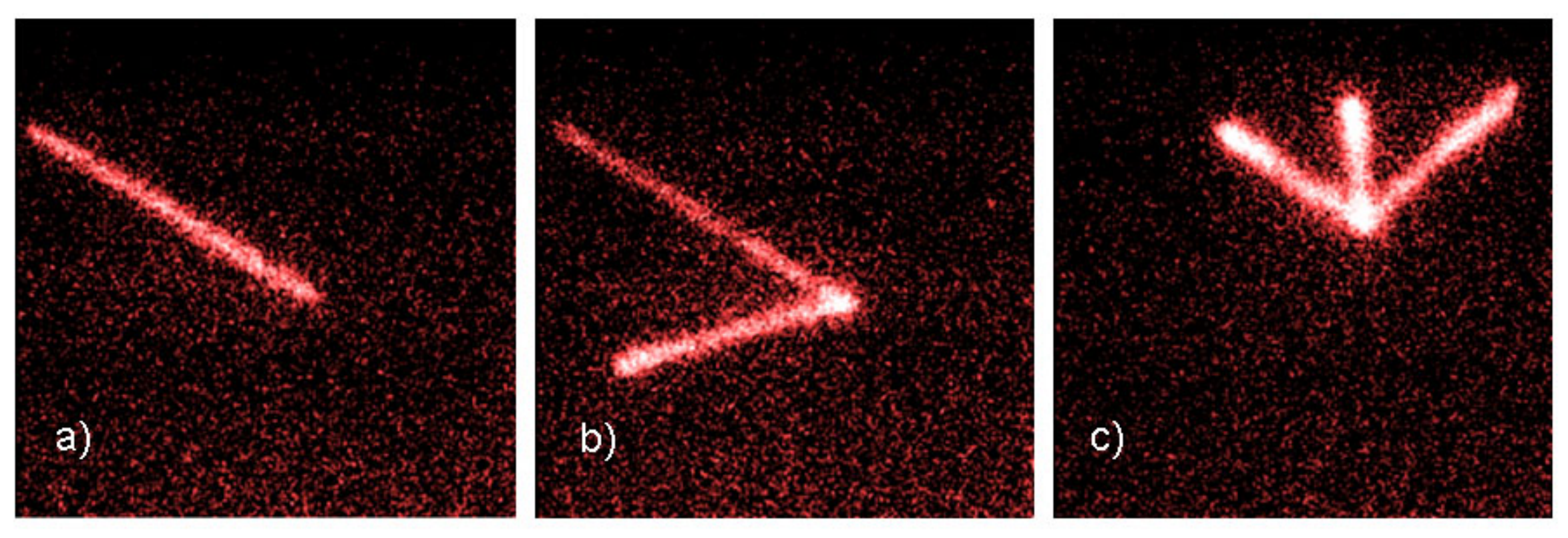}
\caption[T]{
Snapshots of $\beta$-delayed particle emission from $^{43}\mathrm{Cr}$ \cite{Pomorski2011_2}: (a) proton emission, (b) $2p$ emission, and (c) three-proton emission. See Ref.\,\cite{Pomorski2011_2} for details.
\label{Pomorski2011}}
\end{figure*}

{\it Time projection chamber.} --- The opening angle and relative momentum correlations can be reconstructed from the signals measured in the experiment via the two abovementioned methods; however, the physical process of decay cannot be directly observed. Time projection chambers (TPCs) are an effective detection technology that can intuitively ``see’’ the decay process and particle tracks. TPCs are a type of particle track detector widely used in nuclear and particle physics to measure the three-dimensional trajectory and energy information of particles \cite{XuJY2018}. For the decay from proton-rich nuclei (e.g., $^{45}\mathrm{Fe}$), the traces of two valence protons are measured using an optical time projection chamber-charge coupled device (OTPC-CCD) method \cite{Miernik2009}. A sketch of the OTPC-CCD detector setup is shown in Fig.\,\ref{Miernik2009}. The entire geometric volume of the OTPC is $20\times20\times15$\,cm$^3$; this is filled with a gas mixture of 49\% He, 49\% Ar, 1\% N$_{2}$, and 1\% CH$_{4}$ at atmospheric pressure. The protons emitted from proton-rich nuclei typically have an energy ranging from a few hundred keV to a few MeV. They can be absorbed in gas-detector-based TPCs, and they exhibit clear tracks. For $^{45}\mathrm{Fe}$, the two protons emitted in the $2p$ decay process share an energy of 1.15\,MeV. In the most probable case (i.e., of equal energy division), the proton with an energy of ~0.5\,MeV has a sufficiently long track (~2.3\,cm) in the counting gas of the OTPC. However, for a 4\,MeV proton (as emitted in the $\beta$-delayed $2p$ decay of certain proton-rich nuclei), the length of the proton track is ~50\,cm, which means the protons will escape the chamber. Therefore, the lower limit of the proton energy can only be determined via TPC if the proton energy is too high \cite{Miernik2009}. After the incident nucleus decays in the detector, the ultraviolet signals generated by ionization are converted to visible light signals by a wave-length-shifter. Then, these signals are recorded by a CCD camera and a photomultiplier tube (PMT) through a glass window. The sampling frequency of the PMT is 50\,MHz, which can record the time of signal drift and yields the spatial $Z$-direction information of particles. As an example, the results of the $\beta$-delayed $1p$, $2p$, and $3p$ emissions of $^{43}\mathrm{Cr}$, as measured by OTPC-CCD in the experiment at NSCL, are shown in Fig.~\ref{Pomorski2011}; here, the track of the emitted proton can be intuitively seen \cite{Pomorski2011_2}.

The abovementioned three experimental methods for measuring the $2p$ emission of proton-rich nuclei can be adapted to different conditions. The method of in-flight decay is useful for measuring decay processes with a short half-life (e.g., the $2p$ decay from light nuclei). The implantation decay method is suitable for the measurement of long-lived nuclei, especially for $\beta$-delayed decay. TPCs place no clear requirements upon the half-life of decay: both short half-life and long half-life (e.g., $\beta$-delayed) decays can be measured. In particular, the OTPC-CCD can be used to observe decay processes and particle traces. The energy of emitted protons can also be obtained by the OTPC-CCD; however, the precision is generally lower than that measured by silicon detectors. In recent years, detection technology has developed rapidly \cite{Duan2018,Liu2019}. Similar to its development in the field of nuclear electronics, waveform sampling technology is becoming increasingly important in data acquisition for nuclear experiments \cite{Grzywacz2003}. If these new techniques are used in experiments, the original signal of the detector can be completely saved, the data can be analyzed and processed offline, and the applicability of the above three methods can be extended.

\subsection{Identification of 2{\it p} emission mechanism}

\begin{figure*}[htb!]
\includegraphics[width=1\textwidth]{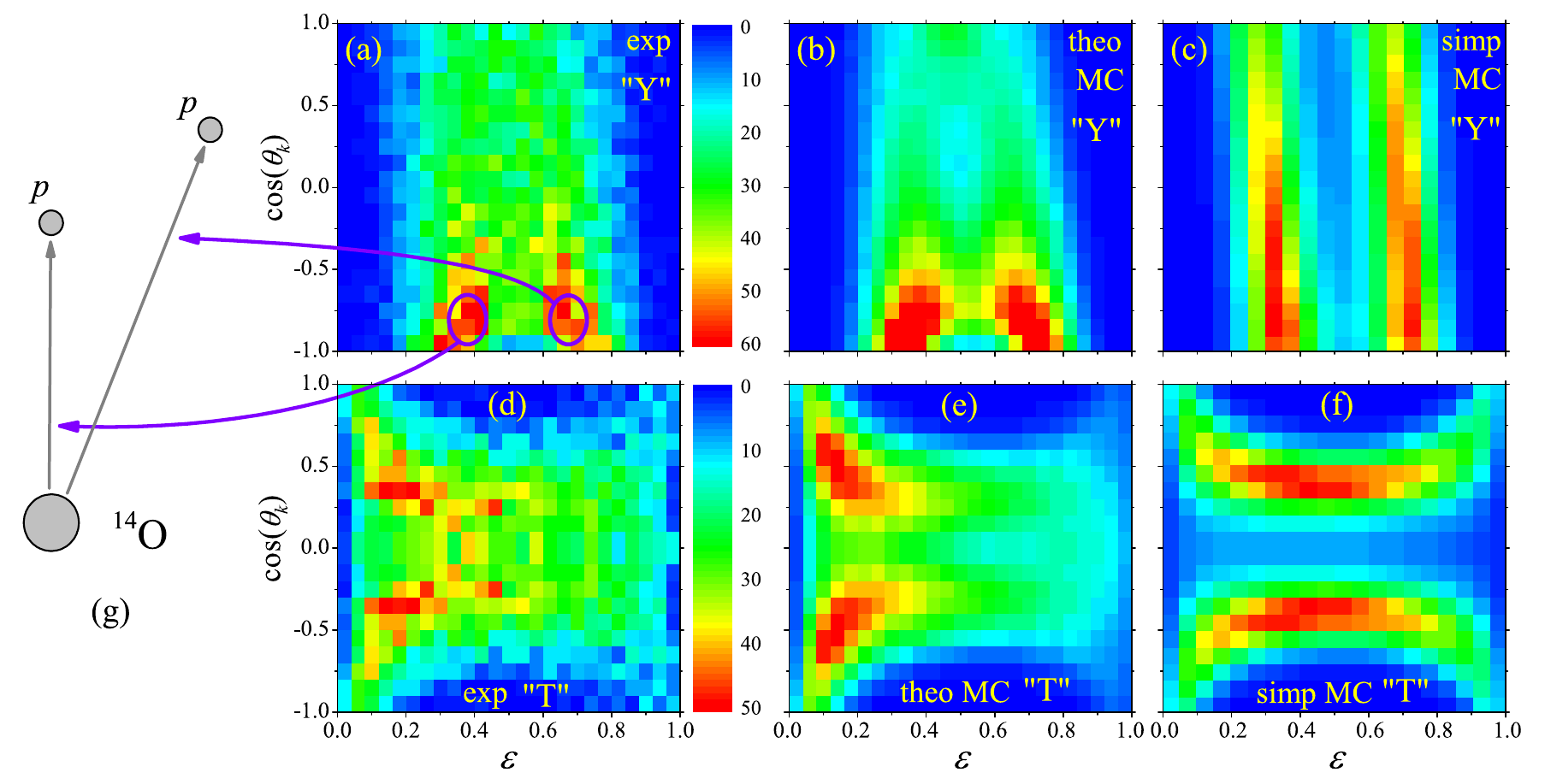}
\caption{Comparison of the experimental Jacobi-Y (a) and -T (d) correlation distributions against those of the three-body model
[(b) and (e)] and those from a sequential decay simulation [(c) and (f)]. The effects of the detector efficiency and resolution upon the theoretical distributions have been included via Monte Carlo simulations. (g) The relative orientations and magnitudes of the two-proton velocity vectors for the peak regions indicated by blue circles in Panel (a). See Ref.\,\cite{Brown2015} for details.
\label{Brown2015}
}
\end{figure*}

As shown in Fig.\,\ref{Fang2020}, three mechanisms are available for $2p$ emission. Typically, the emission time or $pp$ correlation of two emitted protons can differ between these decay processes; this can be used to distinguish the decay mechanisms of $2p$ emissions. For a 2$p$ decay from the ground or low-lying states of the light nuclei, the energy for the relative motions of two valence protons is compatible with the decay energy $Q_{2p}$. Moreover, the Coulomb barrier is relatively small, and the diproton structure can be distorted by Coulomb repulsion after tunneling. As a result, the energy/momentum and angular correlation can be widely distributed, even in the cases exhibiting diproton emission. $^6$Be is one such case; it is thought to exhibit both diproton and large-angle (three-body) emissions compared with the theoretical calculations \cite{Grigorenko2009,Grigorenko2009_2,Oishi2014,Wang2021}; meanwhile, the observed $pp$ correlation has a wide energy distribution \cite{Pfutzner2012,Bochkarev1989,Egorova2012}. Similar situations occur in $^{11,12}$O \cite{Grigorenko2002,Grigorenko2004,Webb2018,Webb2019,Wang2019,Wang2021_2}. 
However, one can still extract useful information from these observables. We take the ground-state $2p$ emitter $^{16}$Ne as an example: Although from the perspective of energy, its first excited ($J^\pi=2^+$) state exhibits sequential decay, the measured correlation pattern shows aspects of both sequential and diproton-like decay (see Fig.\,\ref{Brown2015} and Refs.\,\cite{Brown2014,Brown2015}). This suggests that interference between these processes is responsible for the observed features, which can be described in terms of a ``tethered decay''
mechanism.

For the $2p$ decay produced by the highly-excited or mid-heavy nuclei, the diproton peak in the energy/momentum correlation can be more pronounced. For example, experimental measurements of $^{18}\mathrm{Ne}$ show a clear diproton emission mechanism in the 6.15 MeV excited state \cite{Raciti2008}. In the study of $2p$ emissions for excited states of $^{27,28}\mathrm{P}$ and $^{28,29}\mathrm{S}$, conducted at RIBLL, a diproton emission mechanism was also identified using the same method \cite{Xu2013,Lin2009,Xu2010}. In the following, we present in detail a method for identifying the $2p$ decay mechanism. 

\begin{figure*}[!htb]
\includegraphics[width=0.8\textwidth]{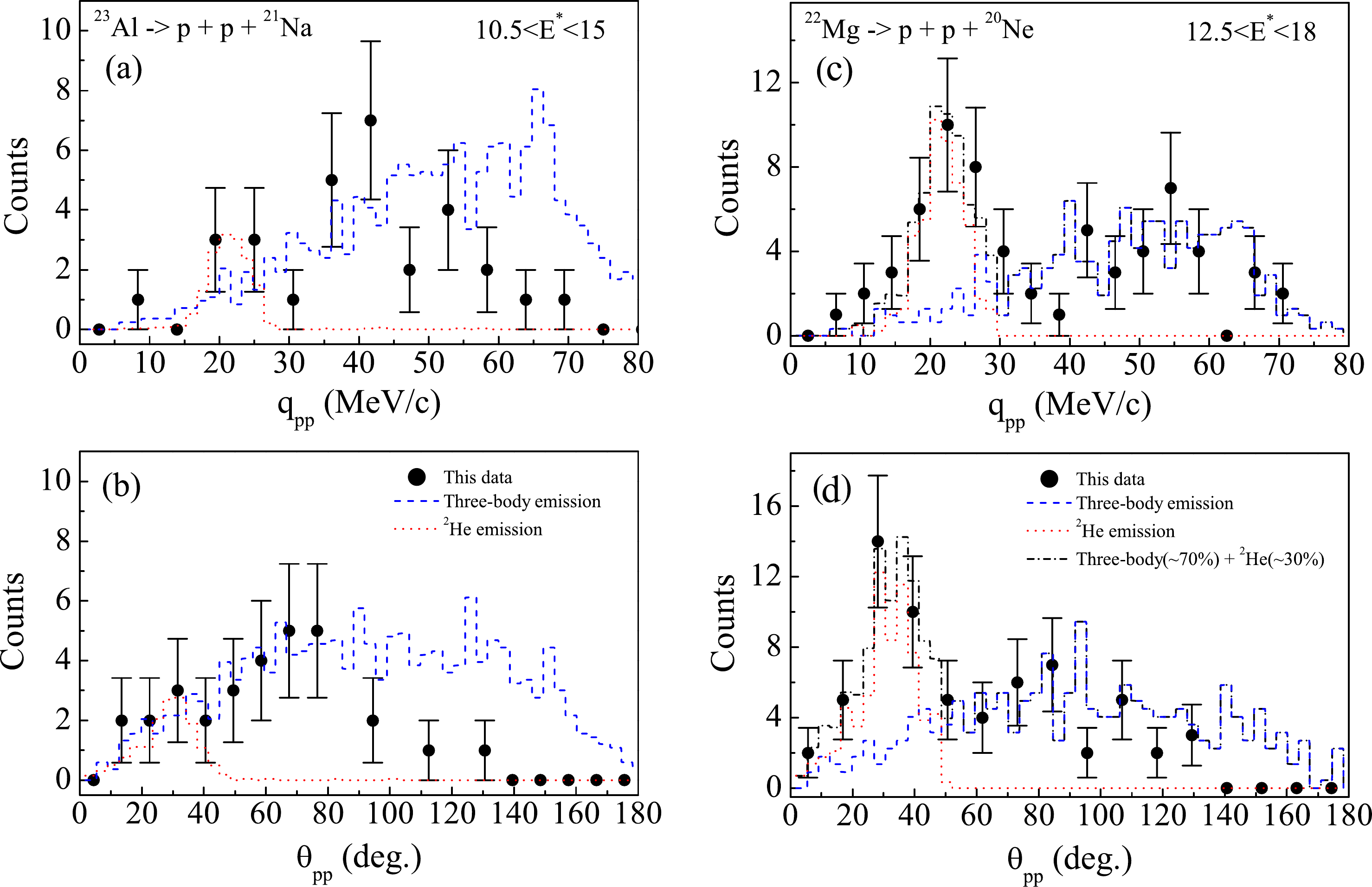}
\caption[T]{
The (a) relative momenta and (b) opening angles between the two protons emitted from the exited states of $^{23}\mathrm{Al}$; the (c) relative momentum and (d) opening angle between the two protons emitted from the exited states of $^{22}\mathrm{Mg}$. See Ref.\,\cite{Ma2015} for details.
\label{Ma2015_2}}
\end{figure*}

As described above, in 1983, a $\beta$-delayed decay experiment performed at the LBNL identified a $2p$ emission in the 14.044 MeV excited state of $^{22}\mathrm{Mg}$; however, the emission mechanism could not be clearly identified because the statistics for the relative angle distribution were too low \cite{Jahn1985}. To elucidate this decay mechanism, the $2p$ emissions of two excited proton-rich nuclei, $^{23}\mathrm{Al}$ and $^{22}\mathrm{Mg}$, were recently measured using the RIPS at the RIKEN RI Beam Factory \cite{Ma2015}. The in-flight decay introduced above was adopted, and a sketch of the detector setup is shown in Fig.~\ref{Ma2015}. The relative momenta and opening angle between the two emitted protons decayed from $^{23}\mathrm{Al}$ and $^{22}\mathrm{Mg}$ and the excitation energies of the decay nuclei were measured. Figure~\ref{Ma2015_2}(a),(b) shows the results obtained by comparing the experimental data and theoretical simulation results for different mechanisms; the highly excited states of $^{23}\mathrm{Al}$ were dominated by the sequential emissions or three-body (large-angle) emissions in different energy ranges. However, as shown in Fig.~\ref{Ma2015_2}(c),(d), around the 14.044 MeV excited state of $^{22}\mathrm{Mg}$, a ~30\% probability that the valence protons would decay as diprotons was observed; the other 70$\%$ were likely to undergo sequential or three-body (large-angle) emission. Meanwhile, the other low-excitation states $^{22}\mathrm{Mg}$ were basically governed by three-body (large-angle) or sequential emission. Consequently, the experiment quantitatively determined the $2p$ decay mechanism for the 14.044\,MeV excited state of $^{22}\mathrm{Mg}$ \cite{Ma2015}; this is useful for further studying the properties of excited $2p$ emissions. 
As a different experimental method, the implantation decay was used to measure the $\beta$-delayed $2p$ emission of $^{22}\mathrm{Al}$ in RIBLL \cite{WangYT2018}; this confirms the diproton emission from the 14.044 MeV excited state of $^{22}\mathrm{Mg}$.
 
\begin{figure}[!htb]
\includegraphics[width=1.0\columnwidth]{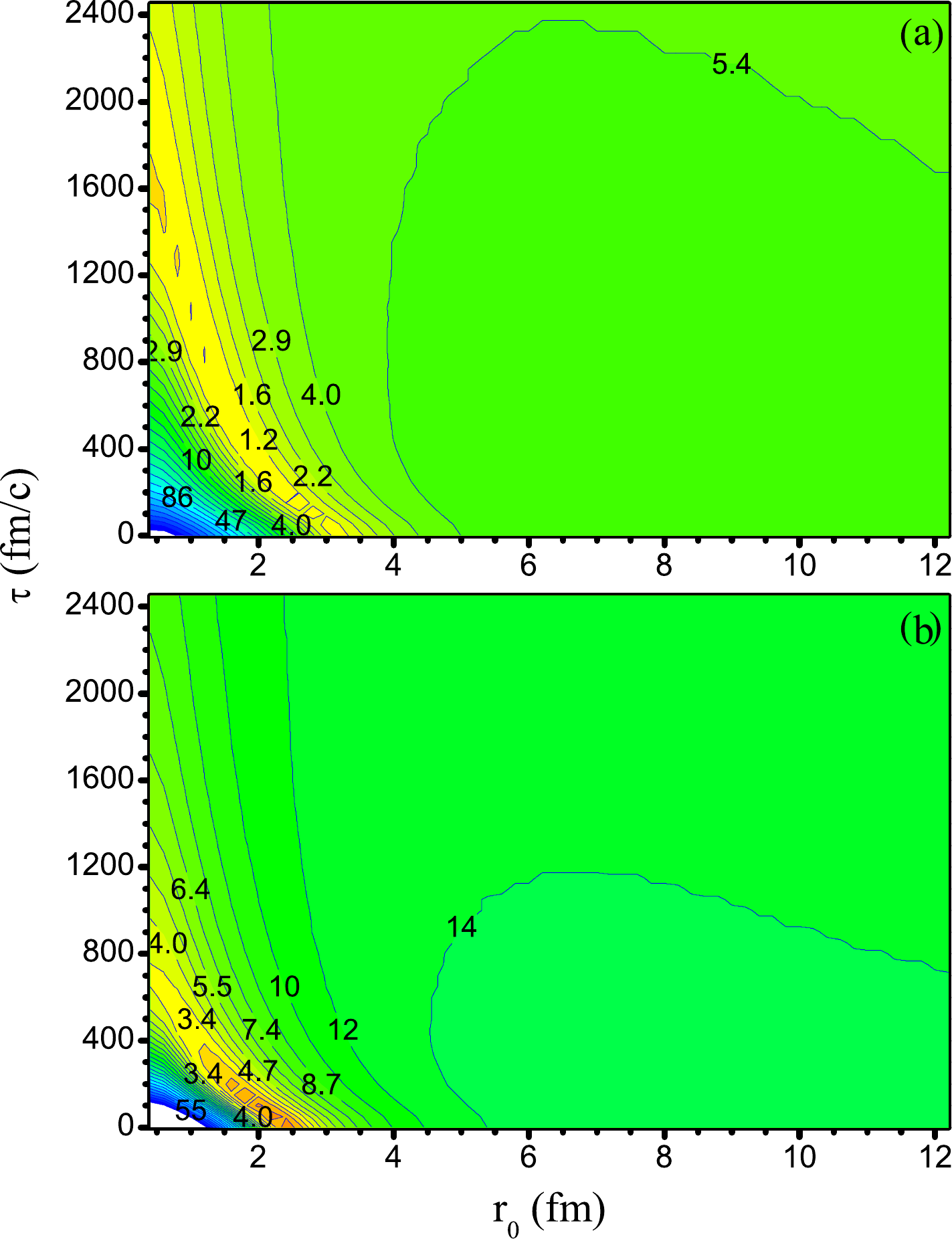}
\caption[T]{
Contour plot of the reduced ${\chi}^{2}$ obtained by fitting the proton--proton momentum correlation function using the CRAB calculation \cite{Fang2016}: (a) $^{23}\mathrm{Al}$ $\rightarrow$ p + p + $^{21}\mathrm{Na}$ and (b) $^{22}\mathrm{Mg}$ $\rightarrow$ p + p + $^{20}\mathrm{Ne}$.
\label{Fang2016}}
\end{figure}

To further analyze the $2p$ decay mechanism, simulation tools are required. The two-particle intensity interferometry (HBT) \cite{Brown1956} method is widely used to obtain the emission source size and emission time for middle- and high-energy nuclear reactions \cite{Koonin1977}. For the simultaneous three-body and sequential emission, the time scales can differ slightly, because the former proceeds via simultaneous emission and the latter exhibits a certain time difference. To clearly distinguish the mechanism of three-body (large-angle) and sequential emission, the HBT method was used in the above study investigating the $2p$ emission of $^{23}\mathrm{Al}$ and $^{22}\mathrm{Mg}$ \cite{Fang2016}. Assuming that the first proton is emitted at time $t_1=0$ and the second proton is emitted at time $t_2=t$, the space and time profile of the Gaussian source can be simulated according to the function $S(r,t)$ $\sim$ $\mathrm{exp}(-{r}^{2}/2{r}_{0}^{2}-t/\tau)$ in the CRAB code \cite{Fang2016}. $r_{0}$ refers to the radius of the Gaussian source, and $\tau$ refers to the lifetime for the emission of the second proton, which starts from the emission time of the first proton. $r$ and $t$ denote the position and time of proton emission, respectively; they were randomly selected. The agreement between the experimental data and CRAB calculations for different $r_{0}$ and $\tau$ values was evaluated by determining the value of the reduced ${\chi}^{2}$, as shown in Fig.~\ref{Fang2016}. The results show that the ranges of the source size parameters were obtained as $r_{0} = 1.2 {\sim} 2.8$\,fm and $r_{0} = 2.2 {\sim} 2.4$\,fm for $^{23}\mathrm{Al}$ and $^{22}\mathrm{Mg}$, respectively. Although the source sizes differ between these two experiments, they have essentially identical effects when compared to the size of the nucleus itself. The ranges of the emission time parameters were obtained as $\tau = 600 {\sim} 2450$\,fm/$c$ and $\tau = 0 {\sim} 50$\,fm/$c$ for $^{23}\mathrm{Al}$ and $^{22}\mathrm{Mg}$, respectively. The large difference in emission time between the two nuclei means that the 2$p$ emission mechanism of $^{23}\mathrm{Al}$ is primarily sequential emission, owing to the long emission-time difference. Meanwhile, the emission mechanism for $^{22}\mathrm{Mg}$ is almost that of three-body emission. Combining this with the relative momenta and opening angles of the two emitted protons (as discussed above), the decay mechanism of $2p$ emission can be clearly identified \cite{Fang2016,Zhou2020}. 

Meanwhile, the HBT method has been used to analyze experimental data from the two-neutron ($2n$) decay, which has also come to represent a frontier of radioactive nuclear beam physics in recent years \cite{Spyrou2012}. The $2n$ emission of $^{18}\mathrm{C}$ and $^{20}\mathrm{O}$ have been measured at GSI, and the measured momentum correlation for the emitted neutrons was analyzed using the HBT method \cite{Revel2018}. The results also indicate that the time scales substantially differ between simultaneous and sequential decay.

So far, only a few ground-state $2p$ emitters have been observed with relatively low statistics; hence, the primary goal is to explore more $2p$ emitter candidates. Meanwhile, the study of the $2p$ decay mechanism and other characteristics has also increased the demand for high-quality experimental data. According to the theory prediction (see the discussions below and Refs.\,\cite{Olsen2013,Olsen2013Sep,Neufcourt2020_2} for details), ground-state $2p$ decay is widespread in the light and mid-heavy nuclei beyond the proton dripline. However, most of these extremely unstable nuclei are difficult to produce via experiments, and the $2p$ emission is sensitive to impacts from other decay channels. The construction of a new generation of high-performance accelerators and the development of novel detection technology [e.g., the Facility for Rare Isotope Beams (FRIB) in the United States, the High Intensity Accelerator Facility (HIAF) in China, and the Rare Isotope Accelerator Complex for On-line Experiment (RAON) in South Korea) will provide better experimental conditions for the further study of radioactivity in exotic nuclei, including $2p$ and $2n$ decay.

\section{Theoretical Developments}

In recent years, many related theoretical developments have been made in various directions. In this section, we briefly introduce certain innovative and important ones. Each focuses upon certain aspects of 2$p$ decays, such as their candidates, inner structures, or asymptotic observables, the impacts of the low-lying continuum, or the interplay between these aspects. These different approaches form a comprehensive description of the 2$p$ decay process, which is useful for better understanding its open quantum nature and exotic properties.

\subsection{Towards quantified prediction of the 2{\it p} landscape}

{\it Exploring the candidate 2$p$ emitters using density functional theory.} --- 
The ground-state 2$p$ decay usually occurs beyond the 2$p$ dripline; here, single-proton emission is energetically forbidden or strongly suppressed owing to the odd-even binding energy staggering attributable to the pairing effect \cite{Pfutzner2012,Blank2008,Blank2008_2,Pfutzner2013,Gao2021,Benzaid2020}. Meanwhile, the presence of the Coulomb barrier has a confining effect upon the nucleonic density; hence, relatively long-lived 2$p$ emitters should be expected. So far, the 2$p$ decay has been experimentally observed in a few light- and medium-mass nuclides with Z $\le$ 36. This raises an urgent demand for theoretical studies to identify further candidates, to better understand the properties of this exotic process \cite{Nazarewicz1996,Ormand1997,Goncalves2017}. To this end, density functional theory (DFT) represents an effective tool offering a microscopic framework and moderate computational costs; it has been widely and successfully applied to describe the bulk properties of nuclei across the entire nuclear landscape \cite{Brown2002,Erler2012,Wu2020}.

\begin{figure}[!htb]
\includegraphics[width=1.0\columnwidth]{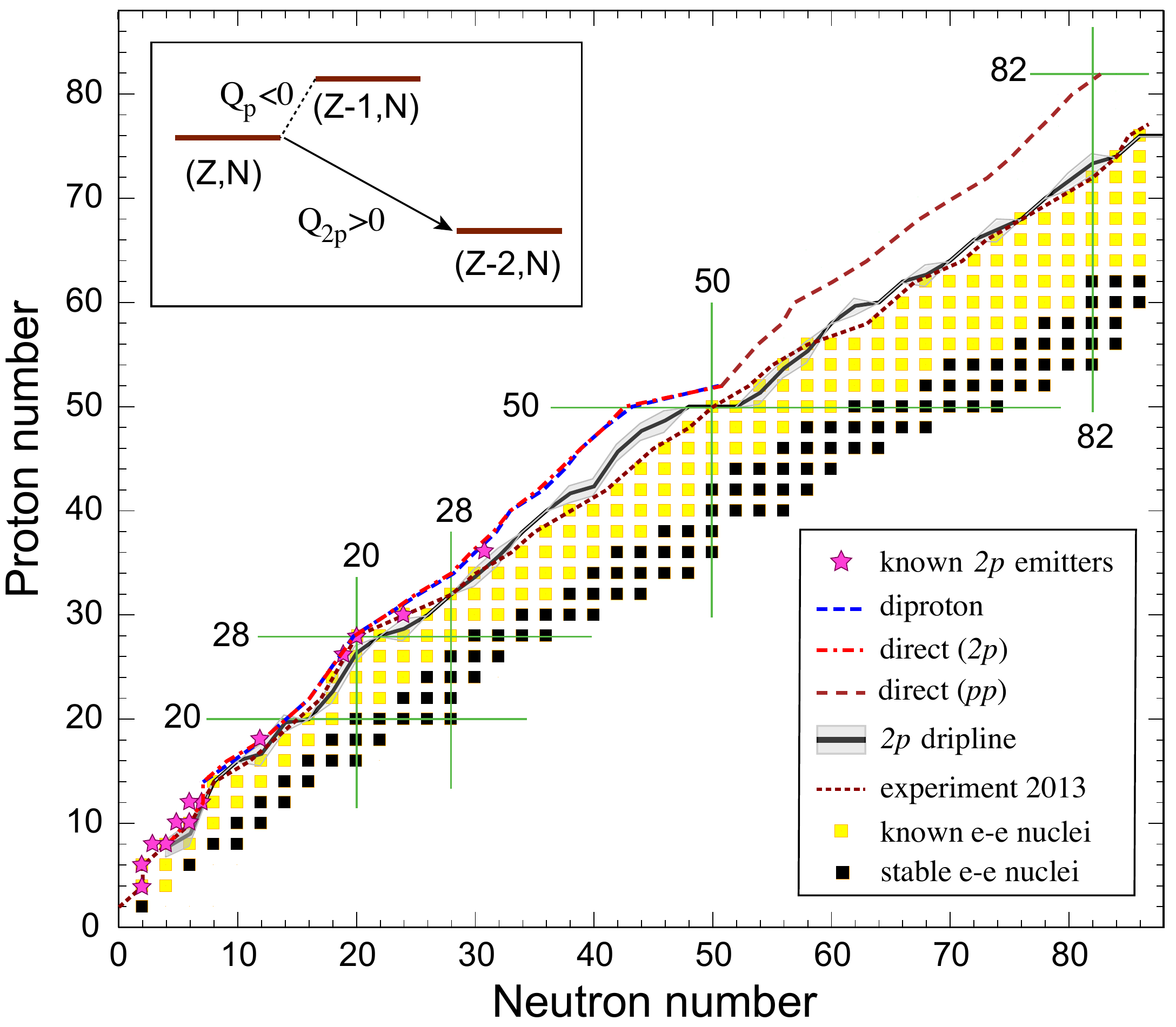}
\caption[T]{The landscape of ground-state $2p$ emitters. The mean $2p$ dripline
(thick black line) and its uncertainty (grey) were obtained in Ref.~\cite{Erler2012} by averaging the
results of six interaction models. The known proton-rich even-even nuclei are denoted by yellow squares, stable even-even nuclei are denoted by black squares, and parts of the known $2p$ emitters are denoted by stars. The current experimental reach for even-$Z$ nuclei (including odd-$A$ systems) \cite{Thoennessen2016} is indicated by a dotted line.
The average lines $N_{\rm av}(Z)$ of $2p$ emission for the diproton model (dashed line)
and direct-decay model (dash-dotted line) are shown.
The energetic condition for the true $2p$ decay is illustrated in the inset. See Ref.\,\cite{Olsen2013,Olsen2013Sep} for details.
\label{Olsen2013_1}}
\end{figure}

The candidate ground-state 2$p$ emitters have been explored using the self-consistent Hartree--Fock--Bogoliubov (HFB) equations \cite{Olsen2013,*Olsen2013Sep}, in which the binding energies for odd-$N$ isotopes are determined by adding the computed average pairing gaps to the binding energy of the corresponding zero-quasiparticle vacuum obtained by averaging the binding energies of even-even neighbors. Because of the Coulomb effect, all the Skyrme energy density functionals in Ref.\,\cite{Olsen2013,*Olsen2013Sep} produced a very consistent prediction for the $2p$ dripline. The corresponding mean value and uncertainty are shown in Fig.\,\ref{Olsen2013_1}. 

In Ref.\,\cite{Olsen2013,*Olsen2013Sep}, the 2$p$ decay candidates were selected according to the decay energy criteria $Q_{2p}$ > 0 and $Q_{p}$ < 0. In this case, single-proton or sequential decay are supposed to be energetically forbidden for medium-mass nuclides (see the inset of Fig.\,\ref{Olsen2013_1}). Meanwhile, the decay half-lives are estimated via 
both direct \cite{Pfutzner2013} and diproton decay \cite{Brown1991,*Brown1991_2} models and compared with those of alpha decay.

\begin{figure}[!htb]
\includegraphics[width=1.0\columnwidth]{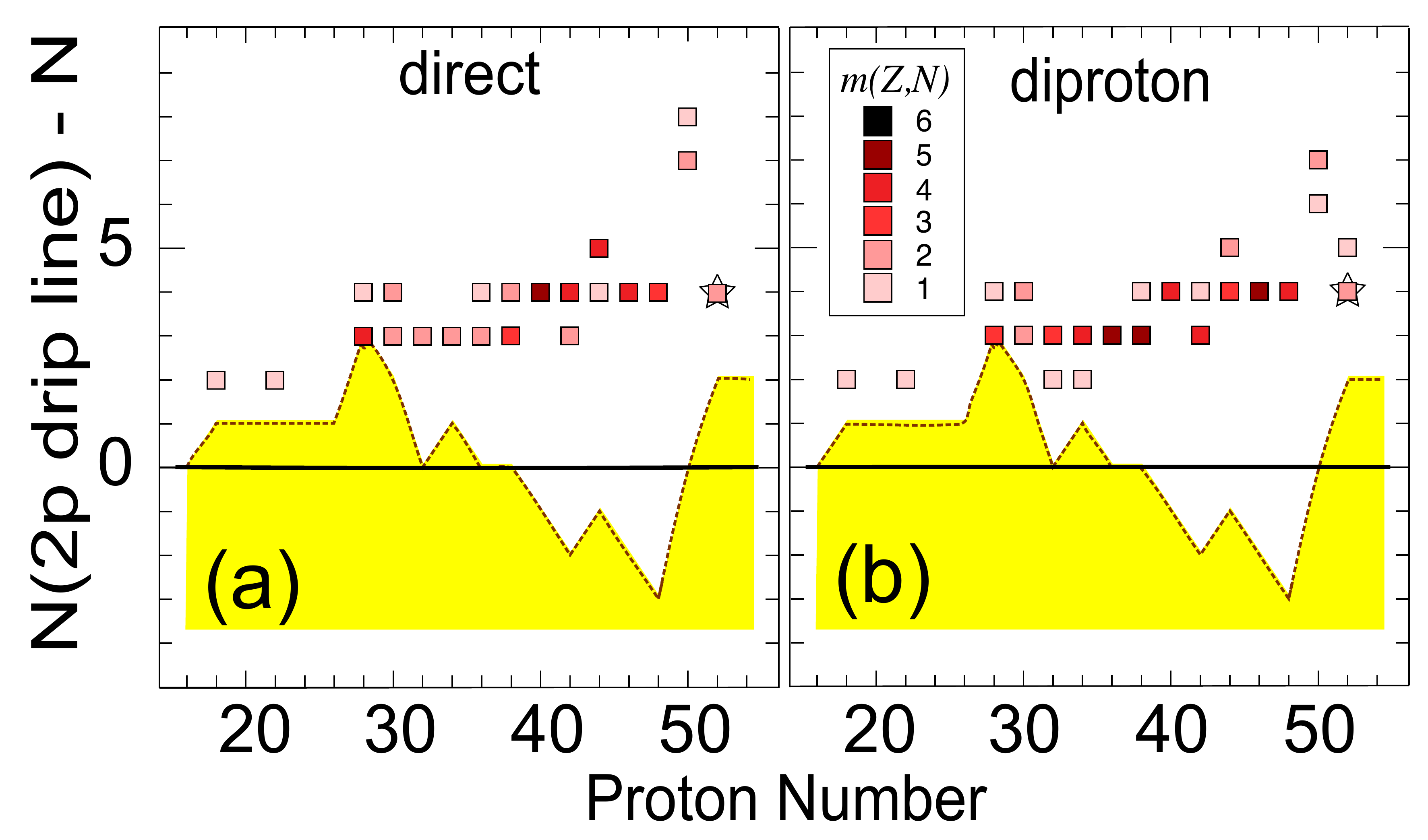}
\caption[T]{The predictions for the (a) direct-decay model and (b) diproton model for the ground-state $2p$ radioactivity. For each value of $Z\ge 18$, neutron numbers $N$ of predicted proton emitters are shown in comparison with the average $2p$ dripline of Ref.~\cite{Erler2012}, as shown in Fig.~\ref{Olsen2013_1}. The model multiplicity
$m(Z,N)$ is indicated by the legend. The candidates for competing $2p$ and $\alpha$ decay are denoted by stars.
The current experimental reach of Fig.~\ref{Olsen2013_1} is denoted by the dotted line. See Ref.\,\cite{Olsen2013,Olsen2013Sep} for details.
\label{Olsen2013_2}}
\end{figure}

As a result, this $2p$ decay mode was shown to not represent an isolated phenomenon but rather a typical feature for proton-unbound isotopes with even atomic numbers. Almost all elements between argon and lead can have $2p$-decaying isotopes (see Fig.\,\ref{Olsen2013_2}); in this range, two regions are most promising for experimental observation: one extends from germanium to krypton and the other is located just above tin. For those nuclei with $Z \le 82$, $\alpha$ decay begins to dominate.

{\it Prediction with uncertainty quantification.} --- To identify the uncertainty in the different theoretical models, and to provide quantitative predictions for further experimental studies, Bayesian methodology has been adopted to combine several Skyrme and Gogny energy density functionals \cite{Neufcourt2020_2}. In the framework of Ref.\,\cite{Neufcourt2020_2}, predictions were obtained for each model $\mathcal{M}_k$ using Bayesian Gaussian processes trained upon separation-energy residuals with respect to the experimental data \cite{Audi2003,WangM2017} and combined via Bayesian model averaging. The corresponding posterior weights conditional to the data $y$ are given by \cite{Hoeting1999,Wasserman2000,Bernardo1994} 
\begin{equation}
p(\mathcal{M}_k|y)
= \frac{p(y|\mathcal{M}_k)\pi(\mathcal{M}_k)}{\sum_{\ell=1}^K p(y|\mathcal{M}_\ell) \pi(\mathcal{M}_\ell)}, 
\end{equation}
where $\pi(\mathcal{M}_k)$ are the prior weights and  $p(y|\mathcal{M}_k)$ denotes the evidence (integrals) obtained by integrating the likelihood over the parameter space.

\begin{figure}[htb!]
\includegraphics[width=1.0\linewidth]{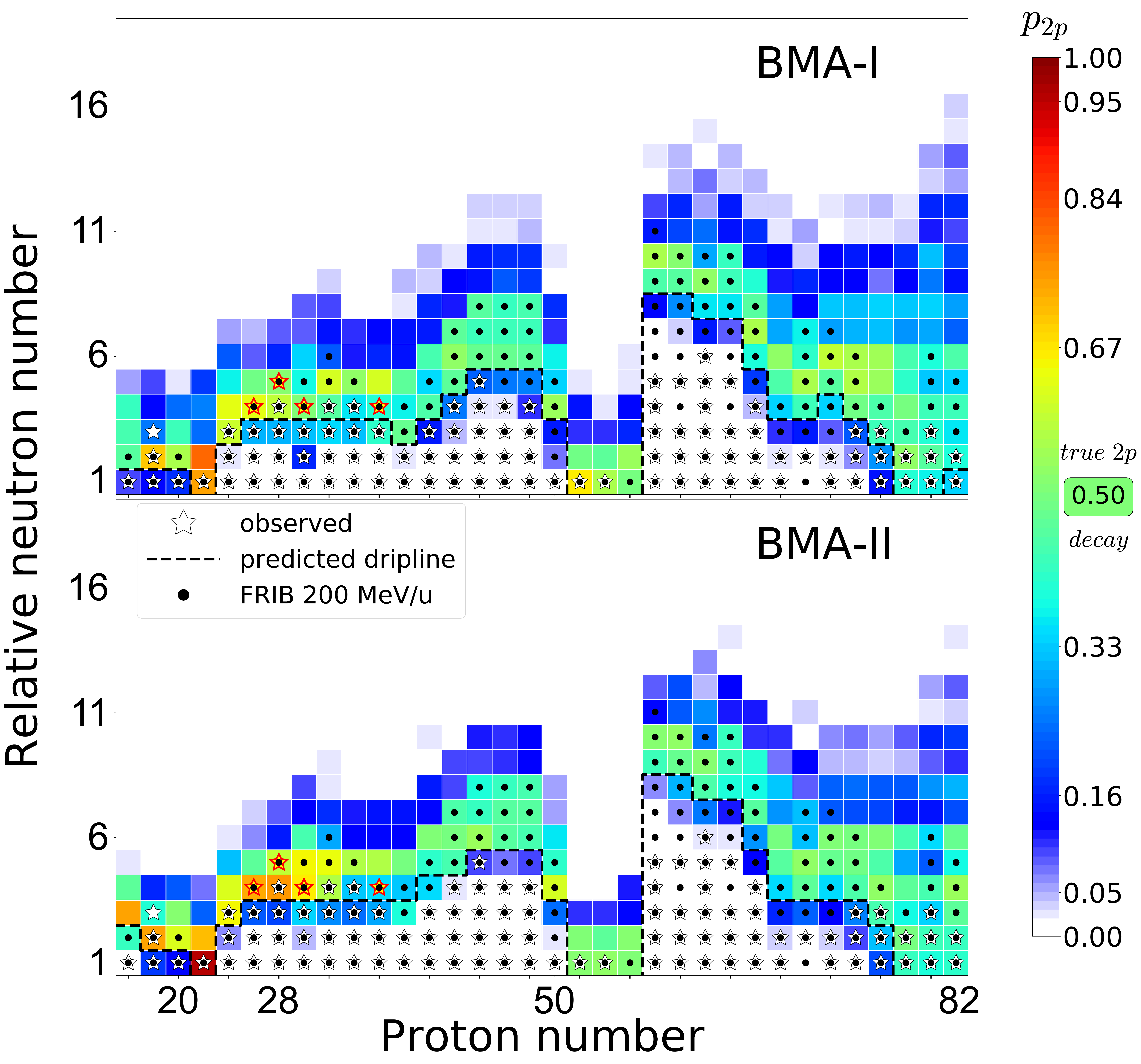}
\caption{Probability of true $2p$ emission for even-$Z$ proton-rich isotopes. BMA-I and BMA-II are the two strategies used to determine model weights \cite{Neufcourt2020_2}.
The color indicates the posterior probability of $2p$ emission (i.e., given that $Q_{2p} > 0$ and $Q_{1p} < 0$) according to the posterior average models.
For each proton number, the relative neutron number $N_0(Z)-N$ is shown, where $N_0(Z)$ is the neutron number of the lightest proton-bound isotope for which an experimental $2p$ separation energy value is available.
The dotted line denotes the predicted dripline (corresponding to $p_{ex} = 0.5$).
The observed nuclei are marked by stars ($^{45}$Fe, $^{48}$Ni,  $^{54}$Zn,
and $^{67}$Kr are denoted by closed stars); those
within FRIB's experimental reach are denoted by dots. See Ref.\,\cite{Neufcourt2020_2} for details.
\label{Neufcourt2020_2}
}
\end{figure}

Within this methodology, a good agreement was obtained between the averaged predictions of statistically corrected models and experiments \cite{Neufcourt2020}. In particular, for this $2p$ decay topic, Ref.\,\cite{Neufcourt2020_2} provides the quantified model results for  $1p$ and $2p$ separation energies and the derived probabilities for proton emission (see Fig.\,\ref{Neufcourt2020_2}); this indicates promising candidates for $2p$ decay and will be tested using experimental data from rare-isotope facilities.

\subsection{Towards a comprehensive description of three-body decay}

\begin{figure*}[htb!]
\includegraphics[width=0.9\linewidth]{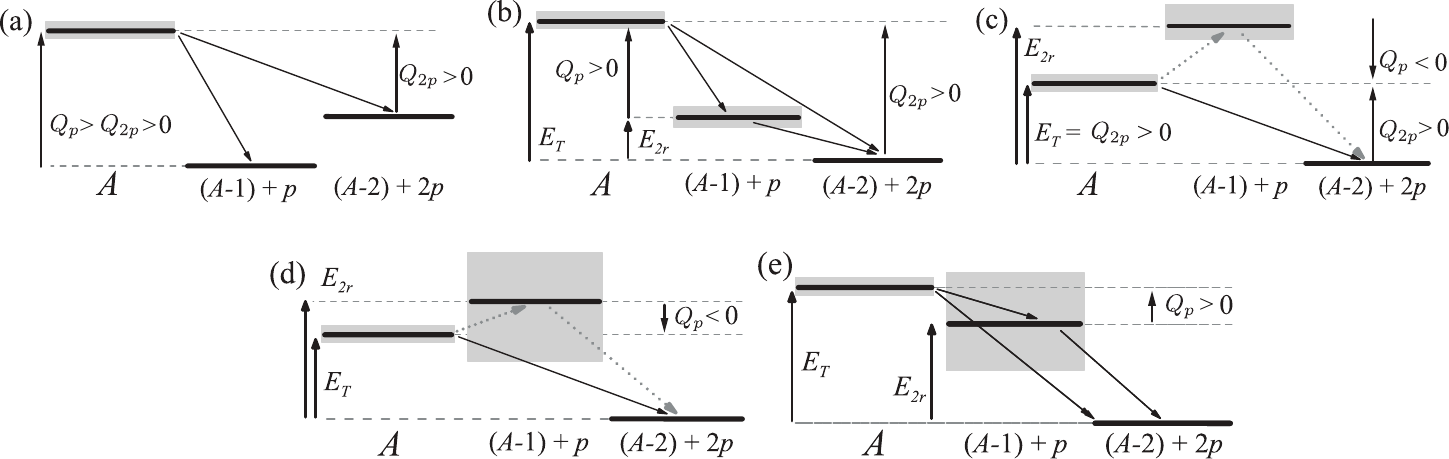}
\caption{Energy conditions for different modes of 
$2p$ emission: (a) typical situation for decays of excited states (both
$1p$ and $2p$ decays are possible), (b) sequential decay via narrow intermediate
resonance, and (c) true $2p$ decays. Cases (d) and (e) indicate ``democratic''
decays, in which the neighboring nucleus has a relatively large decay width. The gray dotted arrows in (c) and (d) indicate the ``decay path''
through the states available only as virtual excitations. See Ref.\,\cite{Pfutzner2012} for details.
\label{Pfutzner2012}
}
\end{figure*}

One of the most fundamental problems of 2$p$ emission is the decay mechanism. In contrast to sequential decay, in which the radioactive process occurs via the intermediate state of the neighboring nucleus, a true (direct) 2$p$ decay occurs when two protons are emitted simultaneously from the mother nucleus. As shown above, to determine whether a nucleus with charge number $Z$ and mass number $A$ produces a true 2$p$ decay, one commonly used selection method (also introduced by Goldansky \cite{Goldansky1960}) is to use the decay energy criteria $Q_{2p}$ > 0 and $Q_{p}$ < 0. This method is effective because the only information needed is the nuclear binding energies. However, the situation is considerably more complicated for realistic cases, as discussed in Ref.\,\cite{Baz1969,Blank2008,Grigorenko2009_2,Pfutzner2012} (see Fig.\,\ref{Pfutzner2012}).

The energy criterion might not be very strict for $2p$ decays originating from the excited states or in the light-mass region. For the former case [as shown in Fig.\,\ref{Pfutzner2012}(a)], the decay energy is typically quite large; in such instances, numerous decay channels are permitted and the decay mechanism can only be roughly estimated using the theoretical structure information or angular momentum transition. For the latter case shown in Fig.\,\ref{Pfutzner2012}(d,e), the mother and neighboring nuclei may have large decay widths owing to the small Coulomb barrier [see Fig.\,\ref{Pfutzner2012}(d,e)]. Consequently, both direct and sequential decay processes are allowed; this is often referred to as ``democratic decay'' \cite{Pfutzner2012,Bochkarev1989}. $^6$Be and the recently discovered $^{11,12}$O are two such cases \cite{Papka2010,Webb2019,Webb2019_2}. 
Moreover, it has been argued that, even though the single-proton decay is energetically forbidden, the mother nuclei can be still be influenced by the intermediate state of the neighboring nuclei via a virtual excitation \cite{Pfutzner2012}.

All of these results indicate that the $2p$ emitter strongly depends on the properties of its neighboring and daughter nuclei, and each one must be studied carefully to provide a comprehensive and accurate description. Those direct or diproton models (which treat two valence protons as a pair for simplicity) are useful for systematical calculations but might not be very helpful for understanding the $2p$ mechanism. This will also introduce large uncertainties in the decay property studies. This complicated $2p$ process is a consequence of the interplay between the internal nuclear structure, asymptotic behavior, and continuum; hence, these effects must be precisely treated using theoretical developments.

Moreover, interest in this exotic decay process has been invigorated by proton--proton correlation measurements following the decay of $^{45}$Fe~\cite{Miernik2007}, $^{19}$Mg~\cite{Mukha2008}, and $^{48}$Ni~\cite{Pomorski2014}; these have demonstrated the unique three-body features of the process and (in terms of theory) the prediction sensitivity for the angular momentum decomposition of the $2p$ wave function. The high-quality $2p$ decay data have necessitated the development of comprehensive theoretical approaches that can handle the simultaneous description of structural and reaction aspects of the problem~\cite{Blank2008,Pfutzner2012}.

{\it Configuration interaction.} --- As one of the most successful models in nuclear theory, configuration interaction (CI) (also known as the interacting shell model) is widely applied in spectroscopic studies across the nuclear landscape. Applying CI to study $2p$ decay can help us to better understand the internal structure information and its impact upon the $2p$ decay width.

\begin{figure}[htb!]
\includegraphics[width=1\linewidth]{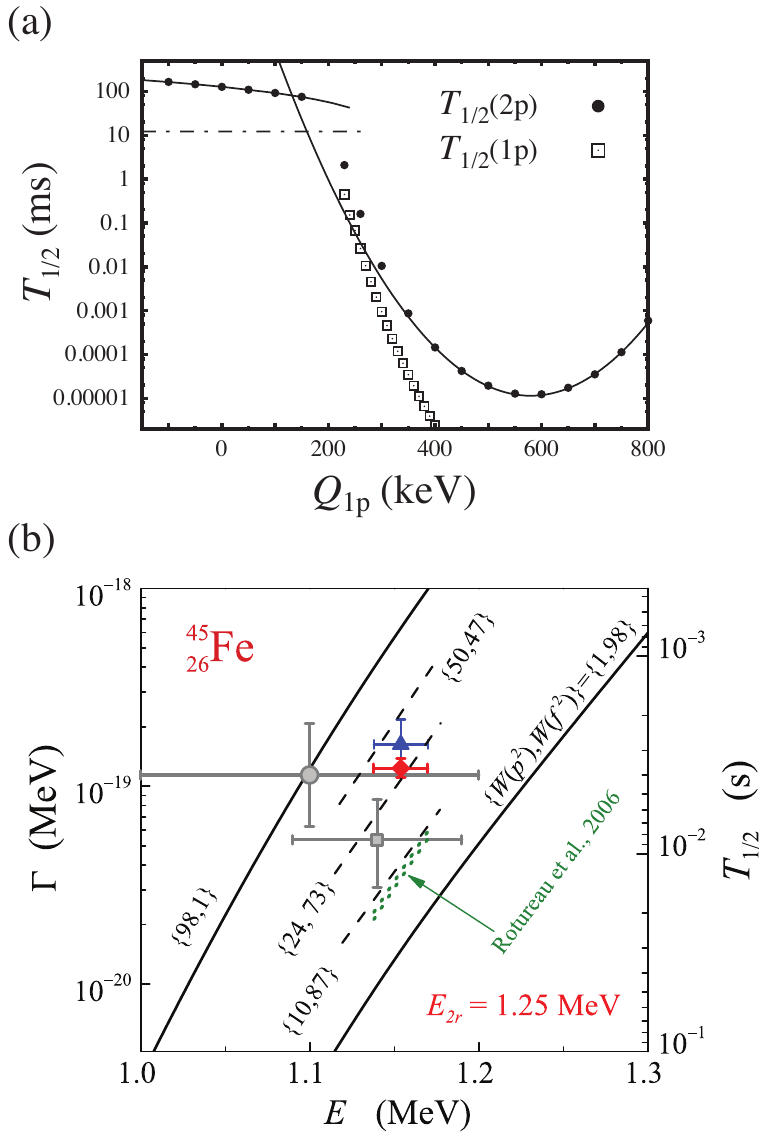}
\caption{(a) The calculated half-life for sequential $2p$ decay from the ground-state $J^\pi$ = 3/2$^+$ in $^{45}$Fe as a function of $Q_{1p}$ (circle), as well as the half-life for $1p$ decay (squares). The dashed-dotted line shows the half-life for the diproton decay. The results were obtained via SMEC calculations. (b) The measured half-life for the ground-state of $^{45}$Fe compared with different theoretical calculations. See Refs.\,\cite{Rotureau2006,Pfutzner2012} for details.
\label{Rotureau2006}
}
\end{figure}

The regular CI model, because of its harmonic-oscillator single-particle basis, cannot capture the continuum effect and three-body asymptotic behaviors in the presence of long-range Coulomb interactions. The former phenomena can be well incorporated into the framework of the shell model embedded in the continuum (SMEC) \cite{Bennaceur2000,Okolowicz2003}, which has been successfully applied to describe the $2p$ decay from the $1^-_2$ state in $^{18}$Ne and other ground-state $2p$ emitters (see Fig.\,\ref{Rotureau2006} and Refs.\,\cite{Rotureau2005,Rotureau2006}).
Regarding the latter phenomena, the contribution of the decay width from the three-body dynamics has been estimated using a hybrid model \cite{Brown2019}.

In this hybrid framework (as shown in Ref.\,\cite{Brown2019}), we can investigate the spectroscopic amplitude via the two-nucleon decay amplitudes (TNAs) for the removal of two protons from the initial state $|A\omega^\prime J^\prime\rangle$, leaving it in the final state $\langle(A-2)\omega J |$ given by the reduced matrix element, as
\begin{equation}
\operatorname{TNA}\left(q_{a}, q_{b}\right)=\frac{\left\langle(A-2) \omega J||\left[\tilde{a}_{q_{a}} \otimes \tilde{a}_{q_{b}}\right]^{J_{o}}|| A \omega^{\prime} J^{\prime}\right\rangle}{\sqrt{\left(1+\delta_{q_{a} q_{b}}\right)(2 J+1)}},
\end{equation}
where $\tilde{a}_{q}$ is an operator that destroys a proton in the orbital $q$, $q$ denotes the ($n$,$\ell$, $j$) quantum numbers of the transferred proton, and $\omega$ contains other quantum numbers. 
For simplicity, only the case with $J^\prime$ = $J$ = $J_o$ = 0 and $q_a$ = $q_b$ = $q$ has been considered. Notably, this spin-singlet nucleon pair might not be spatially close. 
The corresponding $2p$ decay width is estimated in two extreme conditions. 
One assumes that the decay for each orbital $q$ is not correlated with the others. Consequently, the $2p$ decay width can be calculated from an incoherent sum:
\begin{equation}
\Gamma_{i}\left(Q_{2 p}\right)=\sum_{q} \Gamma_{s}\left(Q_{2 p}, \ell^{2}\right)\left[\mathrm{TNA}\left(q^{2}\right)\right]^{2},
\end{equation}
where $\Gamma_{s}$ is the partial decay width for a single channel. The other extreme situation is where all amplitudes combine coherently as in two-nucleon transfer reactions. This gives
\begin{equation}
\Gamma_{c}\left(Q_{2 p}\right)=\left[\sum_{q} \sqrt{\Gamma_{s}\left(Q_{2 p}, \ell^{2}\right)} \mathrm{TNA}\left(q^{2}\right)\right]^2,
\end{equation}
in which the phase is taken as positive for all terms.

\begin{table*}[!htb]
\caption{The half-lives (in ms, ${ }^{19} \mathrm{Mg}$ in picoseconds) calculated by the hybrid model with the incoherent and coherent sum of the different amplitudes contributing to the emission process. As a comparison, the experimental $2 p$ emission half-lives (in ms, ${ }^{19} \mathrm{Mg}$ in picoseconds) are also listed. For the four heavier nuclei, the calculations are obtained with and without the contributions from the $s^{2}$ configuration. See Ref.\,\cite{Brown2019} for details.}
\begin{ruledtabular}
\begin{tabular}{ l  c  c  c  c  c }
Nucleus & $T_{1/2}$ &  \multicolumn{2}{c}{ $T_{1/2}$ without $s^2$ } &  \multicolumn{2}{c}{ $T_{1/2}$ with $s^2$ }\\
$J^\pi$ & Expt. &  Incoherent  & Coherent &  Incoherent  & Coherent\\
\hline \\[-4pt]
$^{19}$Mg 1/2$^-$ & 4.0(15) & & & 0.73$^{+1.5}_{-0.17}$ & 0.20$^{+0.40}_{-0.05}$ \\
$^{45}$Fe 3/2$^+$ & 3.6(4) & 20(8) & 6.6(26) & 5.9(24) & 1.8(7) \\
$^{48}$Ni 0$^+$ & 4.1(20) & 5.1(29) & 1.8(11) & 1.3(6) & 0.43(22) \\
$^{54}$Zn 0$^+$ & 1.9(6) & 1.8(8) & 0.9(4) & 1.7(8) & 0.6(3) \\
$^{67}$Kr 3/2$^-$ & 20(11) & 850(390) & 320(140) & 820(380) & 250(110) \\
$^{67}$Kr 1/2$^-$ & 20(11) & 904(420) & 290(130) & 940(430) & 360(160) \\
\end{tabular}
\label{Brown2019}
\end{ruledtabular}
\end{table*}

As shown in Table\,\ref{Brown2019}, the calculated decay widths incorporating three-body dynamics are substantially improved and agree well with the experimental data. Meanwhile, as discussed in Ref.\,\ref{Brown2019}, it is important to include small $s^2$ components in the decay. These components with low-$\ell$ orbitals are crucial for the decay process (owing to their small centrifugal barriers) and are often introduced via the continuum coupling or core/cross-shell excitations. Moreover, the half-lives of  $^{67}$Kr retain a large discrepancy between the experimental data and theoretical predictions \cite{Brown2019}. 
As a recently observed 2p emitter, the half-life of $^{67}$Kr is systematically over-estimated by various theoretical models \cite{Goncalves2017,Grigorenko2003}. It seems that such a short lifetime can only be reproduced by using a large amount of $s$- or $p$-wave components \cite{Brown2019,Grigorenko2017}, which indicates that deformation effects or certain other factors might be missing from the current theoretical or experimental frameworks \cite{Goigoux2016,Brown2019}.

{\it Three-body model.} --- To study the configuration and correlations of valence protons, a theoretical model was developed from a different basis: the three-body nature of $2p$ decay. In these few-body frameworks, a two-particle emitter can be viewed as a three-body system, comprising a core ($c$) representing the daughter nucleus and two emitted nucleons ($n_1$ and $n_2$). The $i$-th cluster ($i=c,n_1,n_2$) contains the position vector  $\vec{r}_i$ and linear momentum $\vec{k}_i$ \cite{Braaten2006,Baye1994,Hagino2005,Hagino2016,Hagino2016_2,Grigorenko2002,Grigorenko2009_2}.

\begin{figure}[!htb]
\includegraphics[width=1\columnwidth]{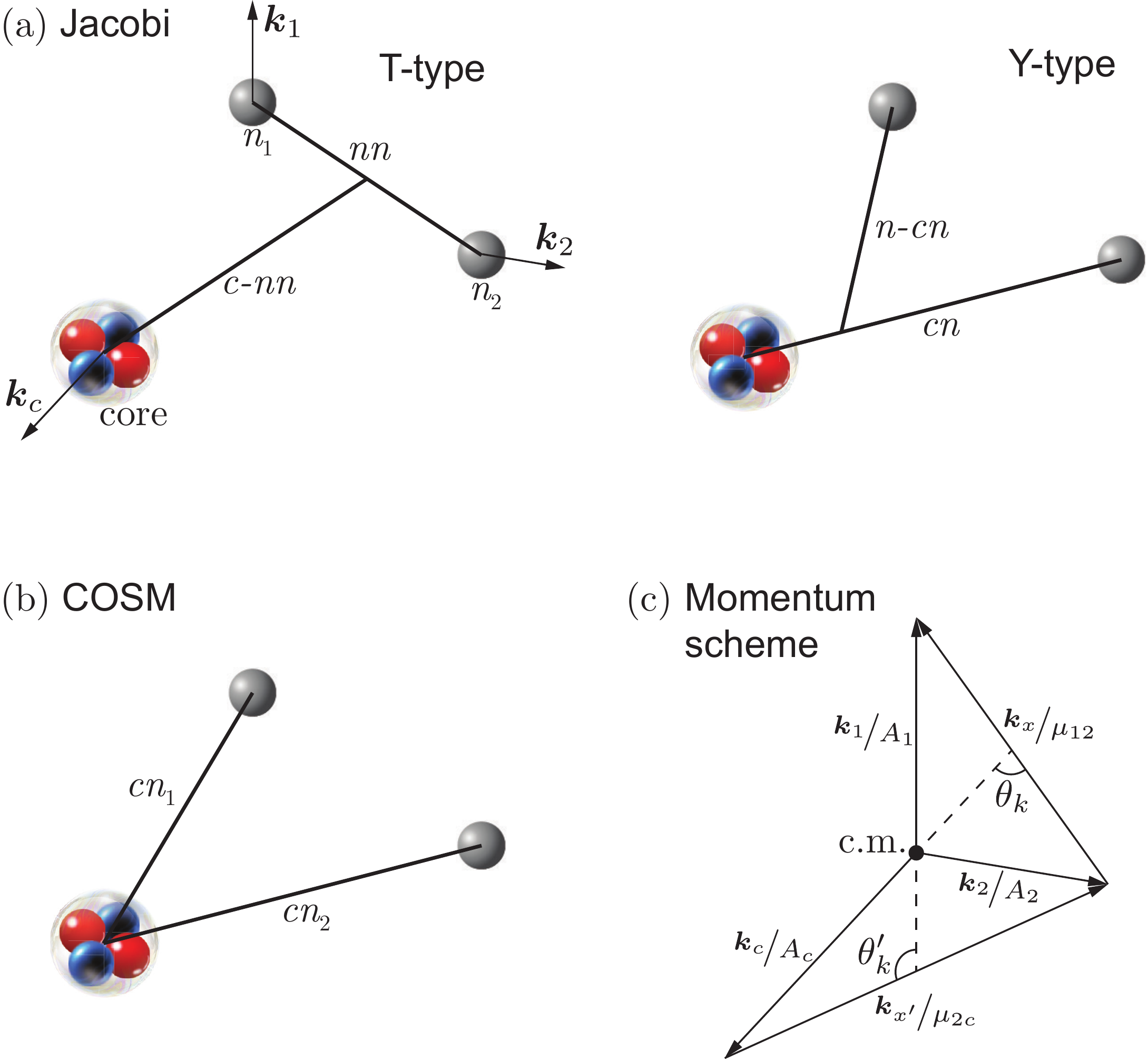}
\caption{Schematic diagram for different coordinates of a core + nucleon + nucleon system: (a) Jacobi T-type, Y-type; (b) cluster-orbital shell model coordinates; and (c) the corresponding momentum scheme. $A$ is the mass number, and $k_1$, $k_2$, and $k_c$ are the momenta of the two nucleons and core, respectively, in the center-of-mass coordinate frame. }\label{Wang2021}
\end{figure}

Two types of coordinates are commonly used to construct the three-body framework. As shown in Fig.\,\ref{Wang2021}, one is the cluster-orbital shell model (COSM)~\cite{Suzuki1988} and the other is Jacobi coordinates~\cite{Grigorenko2002}. The former uses single-particle coordinates ($r_n - r_c$) with a recoil term to handle the extra energy introduced by center-of-mass (c.m.) motion~\cite{Suzuki1988,Michel2002}. In this framework, it is easy to calculate matrix elements and extend them to many-body systems. However, because all the coordinates of the valence nucleons are measured with respect to the core, we must be very careful when treating the asymptotic observables. The latter coordinates are also known as the relative coordinates; they are expressed as
\begin{equation}\label{Jacobi}
\begin{aligned}
    \vec{x} &= \sqrt{\mu _{x}} (\vec{r}_{i_1} - \vec{r}_{i_2}),\\
    \vec{y} &= \sqrt{\mu _{y}} \left( \frac{A_{i_1}\vec{r}_{i_1} + A_{i_2}\vec{r}_{i_2}}{A_{i_1} + A_{i_2}} -\vec{r}_{i_3} \right),\\ 
\end{aligned}
\end{equation}
where ${i_1}=n_1, {i_2}=n_2, {i_3}=c$ for T-coordinates and ${i_1}=n_2, {i_2}=c, {i_3}=n_1$ for Y-coordinates, see Fig.\,\ref{Wang2021}. In Eq.~(\ref{Jacobi})
$A_i$ is  the $i$-th cluster mass number, and  $\mu _{x} =  \frac{A_{i_1}A_{i_2}}{A_{i_1}+A_{i_2}}$ and $\mu _{y} = \frac{(A_{i_1}+A_{i_2})A_{i_3}}{A_{i_1}+A_{i_2}+A_{i_3}} $ are the reduced masses associated with $\vec{x}$ and $\vec{y}$, respectively. The Jacobi coordinates automatically eliminate c.m. motion and allow for the exact treatment of the asymptotic wave functions. Hence, it is widely used in nuclear reactions and in the description of other asymptotic properties. However, the complicated coupling/transformation coefficient and anti-symmetrization prevent it from being extended to larger fermionic systems.

The benchmarking between COSM and Jacobi coordinates has been performed in Ref.\,\cite{Wang2017}, in which both weakly bound and unbound systems ($^6$He, $^6$Li, $^6$Be, and $^{26}$O) were investigated using the Berggren ensemble technique \cite{Berggren1968}. Consequently, the COSM-based Gamow shell model (GSM) and Jacobi-based Gamow coupled-channel (GCC) method gave practically identical results for the spectra, decay widths, and coordinate-space angular distributions of weakly bound and resonant nuclear states \cite{Wang2017}. Furthermore, it has been shown that the Jacobi coordinates capture cluster correlations (e.g., dineutron and deuteron-type correlations) more efficiently.

\begin{figure}[!htb]
\includegraphics[width=0.96\columnwidth]{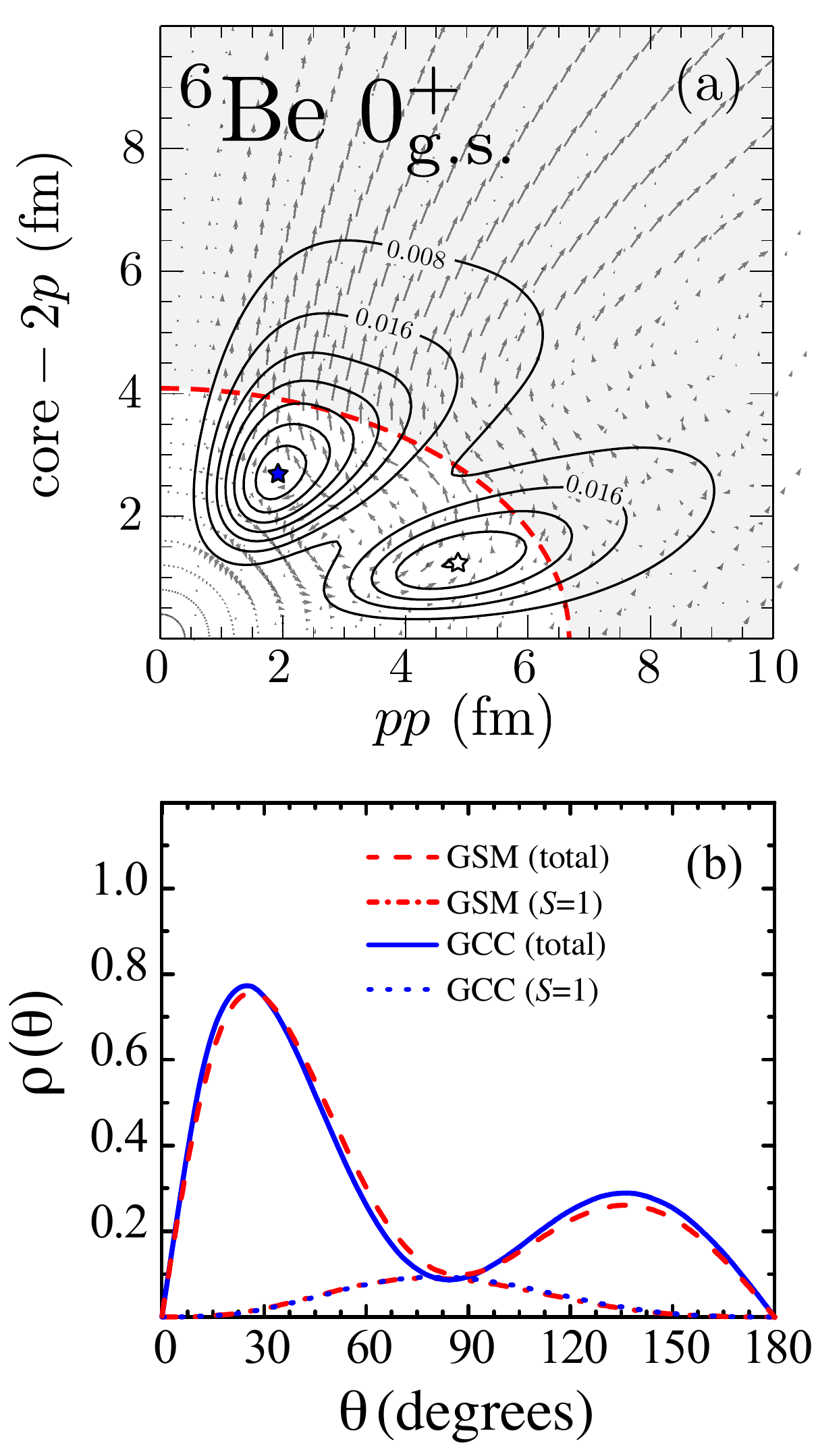}
\caption{(a) Calculated $2p$ density distribution (marked by contours)  and $2p$ flux (denoted by arrows) in the ground state of $^{6}$Be in Jacobi coordinates $pp$ and ${\rm core}-pp$. The thick dashed line marks the inner turning point of the Coulomb-plus-centrifugal barrier. 
The maxima marked by filled and open stars correspond to diproton and cigarlike structures, respectively. (b) Two-nucleon angular densities (total and spin-triplet $S=1$ channels) in the ground-state configurations of $^6$Be, as obtained in GCC and GSM. See Refs.\,\cite{Wang2017,Wang2019} for details.}\label{Wang2019}
\end{figure}

The advantage of the three-body model is that it can capture the configuration of valence protons and the asymptotic correlations of the two-nucleon decay. For example, as a typical light-mass $2p$ emitter, $^6$Be has been considered to feature a ``democratic'' decay mode, attributable to the large width of the ground state of its neighboring nucleus $^5$Li ($Q_p$ = 1.97\,MeV, $\Gamma$ = 1.23\,MeV). The density distribution of $^6$Be has been studied in various few-body models; it shows two maxima associated with diproton and cigarlike configurations (see Fig\,\ref{Wang2019}). This accords with the angular densities obtained by the three-body model \cite{Wang2017,Hagino2005} and GSM \cite{Papadimitriou2011}. This angular density can be calculated using
\begin{equation}
\rho\left( \theta\right)= \int \left\langle\Psi\left|\delta\left(r_{1}-r\right) \delta\left(r_{2}-r^{\prime}\right) \delta\left(\theta_{12}-\theta\right)\right| \Psi\right\rangle {\rm d}r {\rm d}r^\prime,
\end{equation}
which primarily represents the average opening angle of the two valence protons inside the nucleus. Notably, the angular density is defined in coordinate space; it cannot be directly observed in experiments.

Instead, we can look at the configuration evolution of the internal structure during decay by using the flux current $\vec{j} = {\rm Im}({\Psi}^\dagger \nabla \Psi )\hbar/m$, which shows how the two valence protons evolve within a given state wave function $\Psi$.
In the case of $^{6}$Be (as shown in Fig.~\ref{Wang2019}), a competition between diproton and cigarlike configurations occurs inside the inner turning point of the Coulomb-plus-centrifugal barrier associated with the core-proton potential \cite{Wang2019}. Near the origin, the dominant diproton configuration tends to evolve toward the cigarlike configuration, owing to the repulsive Coulomb interaction and the Pauli principle. On the other hand, near the surface, the direction of the flux extends from the cigarlike maximum toward the diproton maximum, tunnelling through the barrier. Moreover, at the peak of the diproton configuration (located near the barrier), the direction of the flux is almost aligned with the core-$2p$ axis, indicating a clear diproton-like decay. Beyond the potential barrier, the two emitted protons tend to gradually separate under the repulsive Coulomb interaction. 
The behavior of the two protons below the barrier can be understood in terms of the influence of pairing, which favors low angular momentum amplitudes; hence, it effectively lowers the centrifugal barrier and increases the probability that the two protons decay by tunnelling~\cite{Wang2017,Grigorenko2009_2,Oishi2014,Oishi2017}.

\begin{figure}[!htb]
\includegraphics[width=0.94\columnwidth]{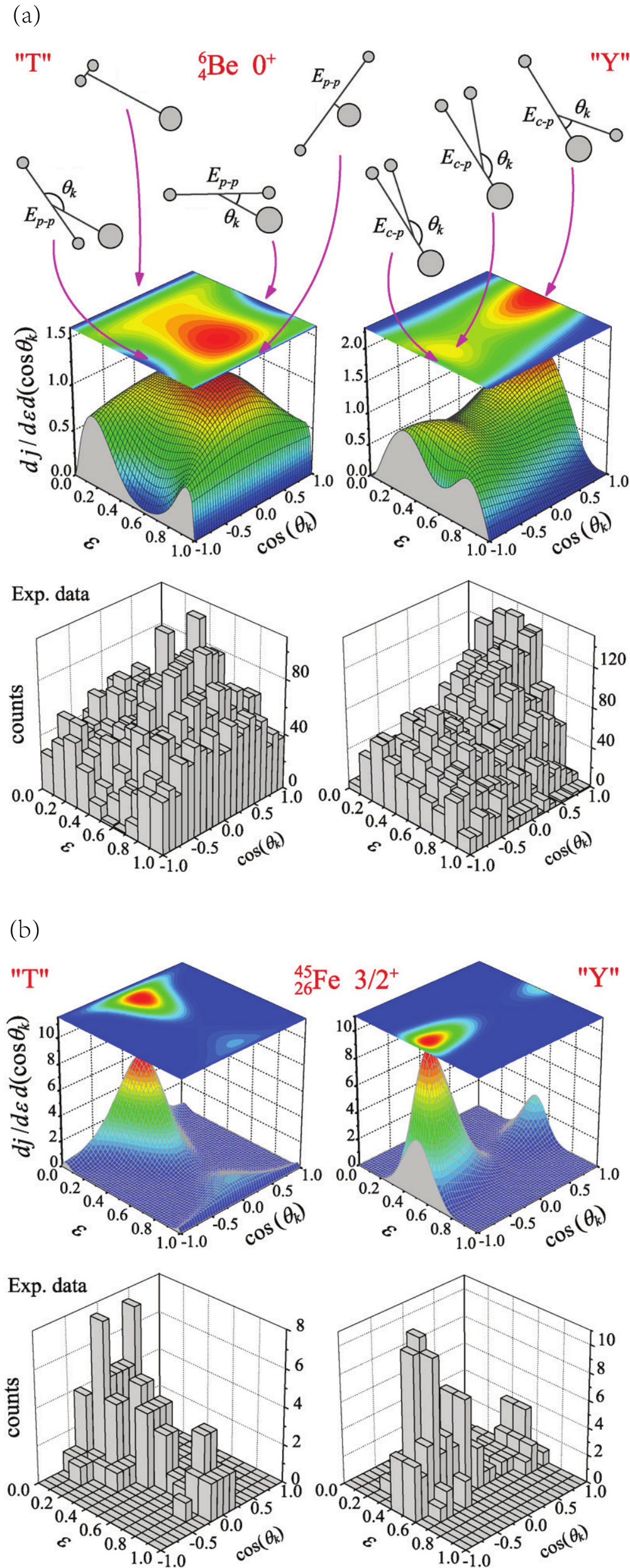}
\caption{Asymptotic nucleon--nucleon correlation for the ground-state $2p$ decay of (a) $^{6}$Be and (b) $^{45}$Fe, represented in Jacobi-T and -Y coordinates (left and right columns, respectively). $\epsilon$ denotes the energy ratio between the subsystem and total $Q_{2p}$, $\theta_k$ is the momentum-space opening angle in Jacobi coordinates. The experimental data are also shown. See Ref.\,\cite{Grigorenko2009} for details.}\label{Grigorenko2009}
\end{figure}

The three-body model also successfully reproduced the asymptotic $pp$ correlations observed in experiments. This nucleon--nucleon correlation indicates the relative energy and opening angle of the emitted protons in the asymptotic region. To elucidate the characteristics of nucleon--nucleon correlation, it is convenient to introduce the relative momenta:
\begin{equation}
\begin{aligned}
    \vec{k}_x &= \mu _{x} \left(\frac{\vec{k}_{i_1}}{A_{i_1}} - \frac{\vec{k}_{i_2}}{A_{i_2}}\right),\\
    \vec{k}_y &= \mu _{y} \left( \frac{\vec{k}_{i_1} + \vec{k}_{i_2}}{A_{i_1} + A_{i_2}} -\frac{\vec{k}_{i_3}}{A_{i_3}} \right).\\ 
\end{aligned}
\end{equation}
No c.m. motion occurs; hence, it is easy to see that $\sum_i \vec{k}_i = 0$, and $\vec{k}_y$ is aligned in the opposite direction to $\vec{k}_{i_3}$. $\theta_k$ and $\theta^\prime_k$ denote the opening angles of ($\vec{k}_x$, $\vec{k}_y$) in Jacobi-T and -Y coordinates, respectively (see Fig.\,\ref{Wang2021}). 
The kinetic energy of the relative motions of the emitted nucleons is given by
$E_{pp/nn} = \frac{\hbar^2 k_x^2}{2 \mu_x}$, and $E_{{\rm core}-p/n}$ denotes that of the core--nucleon pair. Therefore, the $T$-type ($\theta_k$, $E_{pp/nn}$) and $Y$-type ($\theta^\prime_k$, $E_{{\rm core}-p/n}$) distributions reveal the nucleon--nucleon correlation and structural information regarding the mother nucleus. Finally, the total momentum $k$ is defined as $\sqrt{\frac{k_x^2}{\mu_x} + \frac{k_y^2}{\mu_y}}$, which at later times approaches the limit $\frac{\sqrt{2mQ_{2p/2n}}}{\hbar}$, where $Q_{2p/2n}$ is the two-nucleon decay energy given by the binding energy difference between parent and daughter nuclei.

The nucleon--nucleon correlation is important not only for the experimental accessibility but also for the valuable information it carries. For instance, the study of short-range correlations can be used to analyze the fundamental structures of nucleonic pairs at very high relative momenta \cite{Hen2017}. Meanwhile, the long-range correlations of $2p$ decay reveal ground-state properties and the pairing effects of atomic nuclei. Moreover, the decay dynamics and long-range correlations are strongly influenced by both initial-state and final-state interactions \cite{Watson1952,Migdal1955,Phillips1964}; this allows us to gain more insight into the connection between the internal structure and decay properties, including asymptotic correlations.

As shown in Fig.\,\ref{Grigorenko2009}, the asymptotic nucleon--nucleon correlations of $^6$Be, $^{45}$Fe, and several other $2p$ emitters have been reproduced or predicted using the three-body model \cite{Grigorenko2009,Pfutzner2012}. Moreover, from these correlations, one can roughly determine how the valence protons are emitted, as well as the corresponding configurations. For example, in the case of $^{45}$Fe, the small-angle emission dominates in the asymptotic region, which may correspond to a diproton decay. Meanwhile, the situation in  $^{6}$Be is considerably more complicated, with numerous possible configurations.
However, because the emitted protons are influenced by the centrifugal barrier and long-range Coulomb interactions, one must be careful that these configurations properly reflect the circumstances in the asymptotic region, which differ from those inside the nucleus. To better understand how the inner structure evolves into the asymptotic configurations and nucleon--nucleon correlations, several time-dependent frameworks have been developed (see the discussions below and Refs.\,\cite{Oishi2014,Wang2021} for details).

\begin{figure}[!htb]
\includegraphics[width=1\columnwidth]{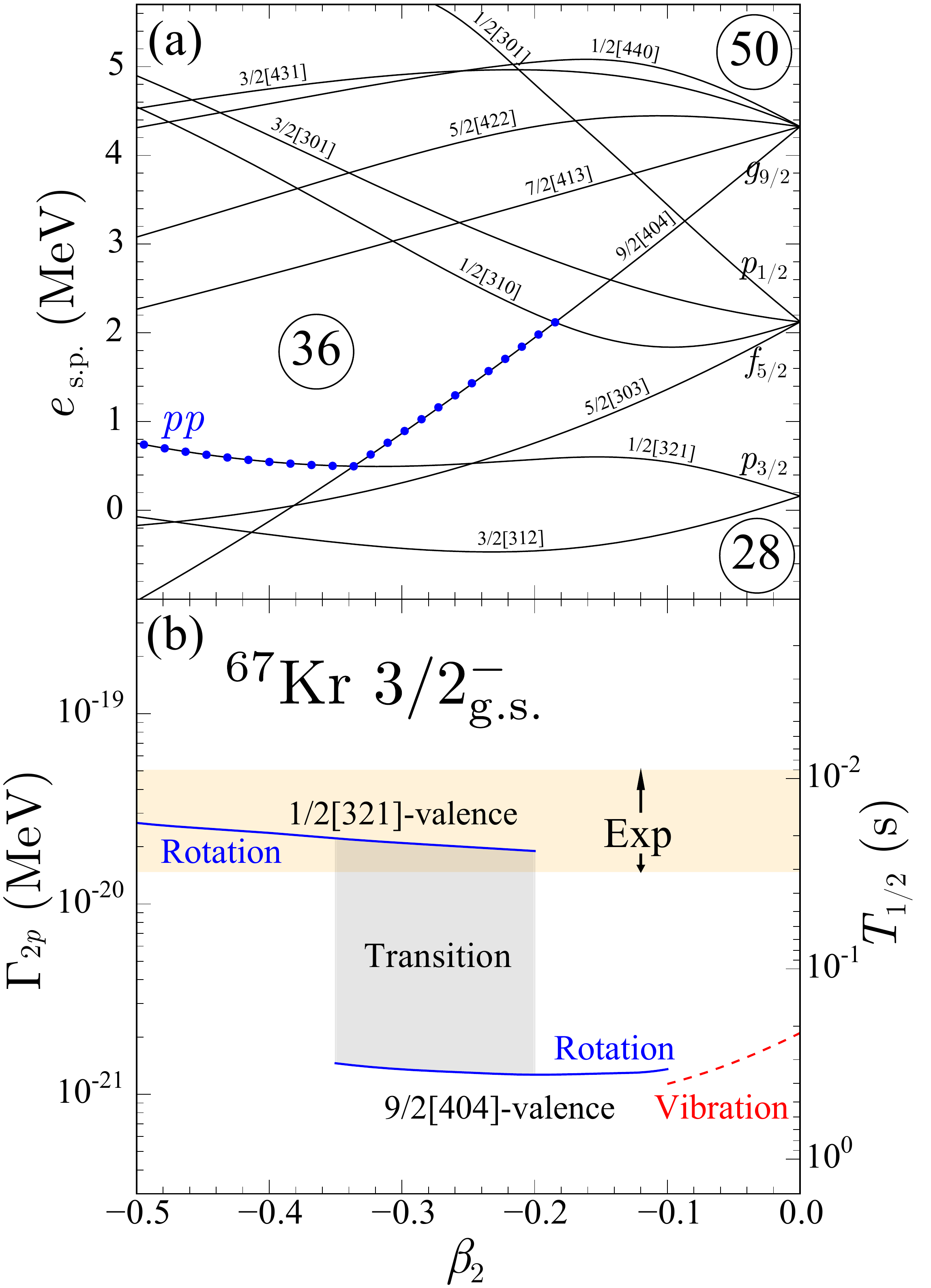}
\caption{Top: Nilsson levels $\Omega$[Nn$_z\Lambda$] of the deformed core-$p$ potential as functions of the oblate quadrupole deformation $\beta_2$ of the core.
The dotted line indicates the valence level primarily occupied by the two valence protons.
Bottom: Decay width (half-life) for  the $2p$ ground-state radioactivity of $^{67}$Kr. The solid and dashed lines denote the results within the rotational and vibrational coupling, respectively. 
The rotational-coupling calculations were performed by
assuming that the 
1/2[321] orbital is either occupied by the core (9/2[404]-valence) or valence (1/2[321]-valence) protons. See Ref.\,\cite{Wang2018} for details.}\label{Wang2018}
\end{figure}

\begin{figure*}[!ht]
     \centering
     \subfigure{\includegraphics[width=\textwidth]{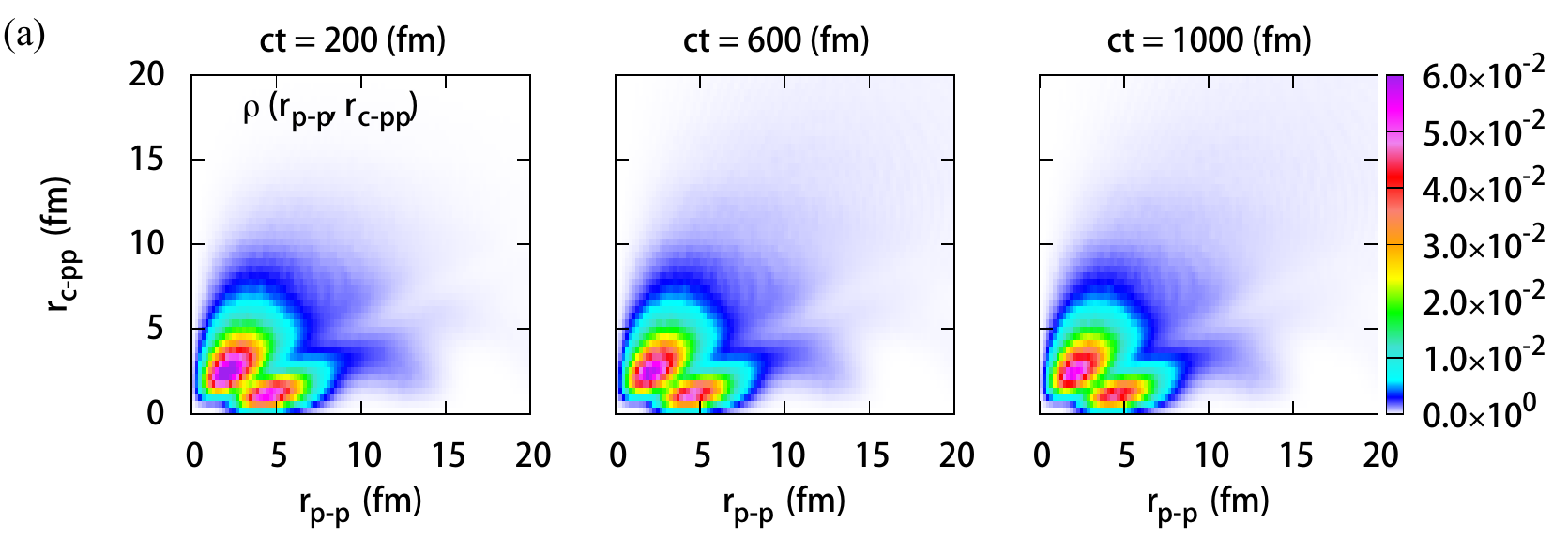}}
     \subfigure{\includegraphics[width=\textwidth]{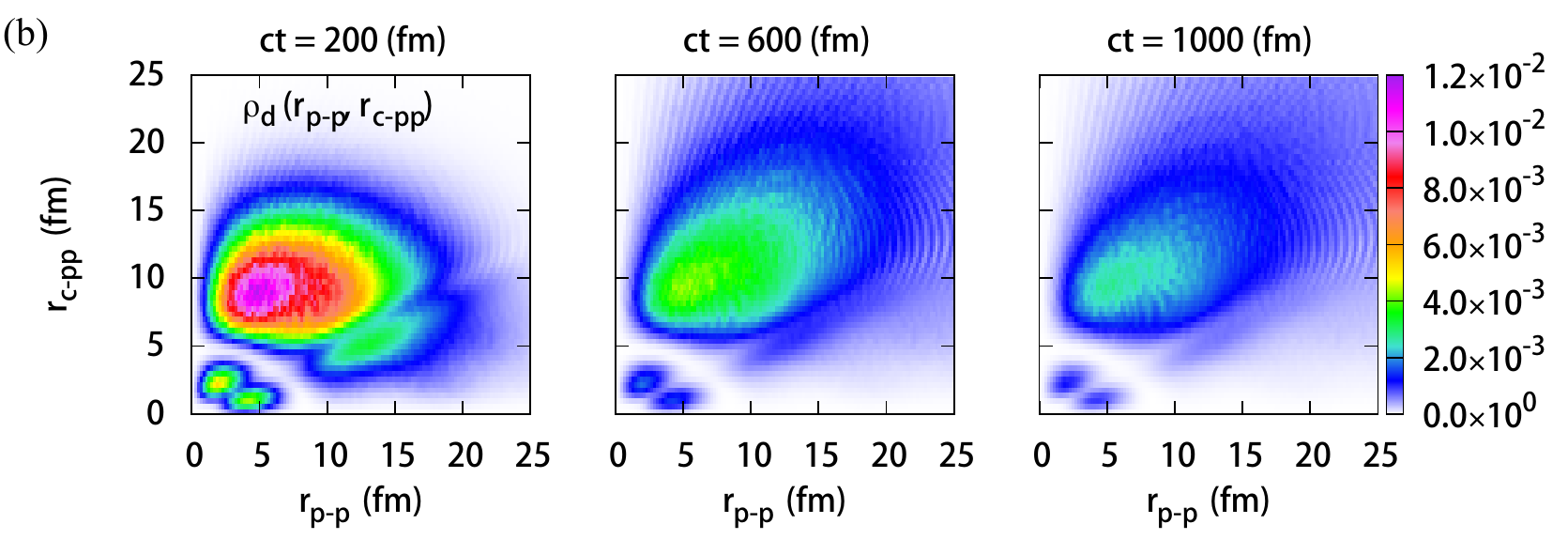}}
        \caption{(a) Time-dependent $2p$ density distribution $\rho$ of the ground state as a function of $r_{p-p}$ and $r_{c-pp}$. (b) The corresponding decaying density distribution $\rho_{d}$. See Ref.\,\cite{Oishi2017} for details.}
        \label{Oishi2017}
\end{figure*}

One main drawback of the regular three-body model is the lack of structural information regarding the daughter nucleus, which is usually treated as a frozen core. To this end, several efforts have been made towards a more microscopic description of the core wave function \cite{Hagino1999,Hagino2001,Barmore2000,Kruppa2004}. 
For $2p$ emission, the GCC method has been 
extended to a deformed case \cite{Wang2017,Wang2018}, which allows a pair of nucleons to couple to the collective states of the core via nonadiabatic coupling. Consequently, the total wave function of the parent nucleus can be written as $\Psi ^{J\pi} = \sum_{J_p \pi_p j_c \pi_c} \left[ \Phi ^{J_p\pi_p} \otimes \phi^{j_c\pi_c} \right]^{J\pi}$, where $\Phi ^{J_p\pi_p}$ and $\phi^{j_c\pi_c}$ are the wave functions of the two valence protons and core, respectively. The wave function of the valence protons $\Phi ^{J_p\pi_p}$ is expressed in Jacobi coordinates and expanded using the Berggren basis~\cite{Berggren1968,Michel2009}, which is defined in the complex-momentum $k$ space. Because the Berggren basis is a complete ensemble that includes bound, Gamow, and scattering states, it provides the correct outgoing asymptotic behavior to describe the $2p$ decay, and it effectively allows nuclear structures and reactions to be treated on the same footing.

As discussed above, the  lifetime of $^{67}$Kr can be influenced by deformation effects \cite{Goigoux2016}. Indeed, studies of single-proton ($1p$) emitters~\cite{Barmore2000,Kruppa2000,Esbensen2000,Davids2001,Davids2004,Kruppa2004,Hagino2001,Florin2003,Arumugam2007} have demonstrated the impact of rotational and vibrational couplings on $1p$ half-lives. The corresponding influence in the $2p$ decay of $^{67}$Kr has been studied using the deformed GCC method \cite{Wang2018}. 
Figure~\ref{Wang2018}(a) shows the proton Nilsson levels (labeled with asymptotic quantum numbers $\Omega$[Nn$_z\Lambda$]) of the Woods--Saxon core-$p$ potential. When the deformation of the core increases, a noticeable oblate gap at $Z=36$ opens up, attributable to the down-sloping 9/2[404] Nilsson level originating from the $0g_{9/2}$ shell. This gap
is responsible for the oblate ground-state shapes of proton-deficient Kr isotopes \cite{Nazarewicz1985,Yamagami2001,Kaneko2004}. The structure of the valence proton orbital changes from the 9/2[404] ($\ell=4$) state at smaller oblate deformations to the 1/2[321] orbital, which has a large $\ell=1$ component. This transition can dramatically change the centrifugal barrier and decay properties of $^{67}$Kr. 
Figure~\ref{Wang2018}(b) shows the $2p$ decay width predicted in the two limits of the rotational model:
(i) the 1/2[321] level belongs to the core, and the valence protons  primarily occupy the 9/2[404] level; (ii) the valence protons  primarily occupy the 1/2[321] level. 
In reality, the core is not rigid; hence, proton pairing is expected to produce a diffused Fermi surface, and the transition from (i) to (ii) becomes gradual, as indicated by the shaded area in Fig.~\ref{Wang2018}(b).  
The decreasing $\ell$ content of the $2p$ wave function results in a dramatic increase in the decay width. At the deformation  $\beta_2 \approx -0.3$, the calculated  $2p$ ground-state half-life of $^{67}$Kr is $24^{+10}_{-7}$\,ms \cite{Wang2018}, which agrees with experimental results~\cite{Goigoux2016}. In future quantitative studies, a more microscopic description of the core might be needed for the three-body model. Furthermore, high-statistics data are expected to improve our understanding of $2p$ radioactivity.

{\it Time-dependent framework.} --- An alternative strategy for tackling the decay process is the time-dependent formalism, which allows a broad range of questions (e.g., configuration evolution \cite{Volya2014}, decay rate \cite{Peshkin2014}, and fission \cite{Bender2020}) to be addressed in a precise and transparent way. In the case of two-nucleon decay, the measured inter-particle correlations can be interpreted in terms of solutions propagating for long periods. 

\begin{figure*}[!htb]
\includegraphics[width=0.75\textwidth]{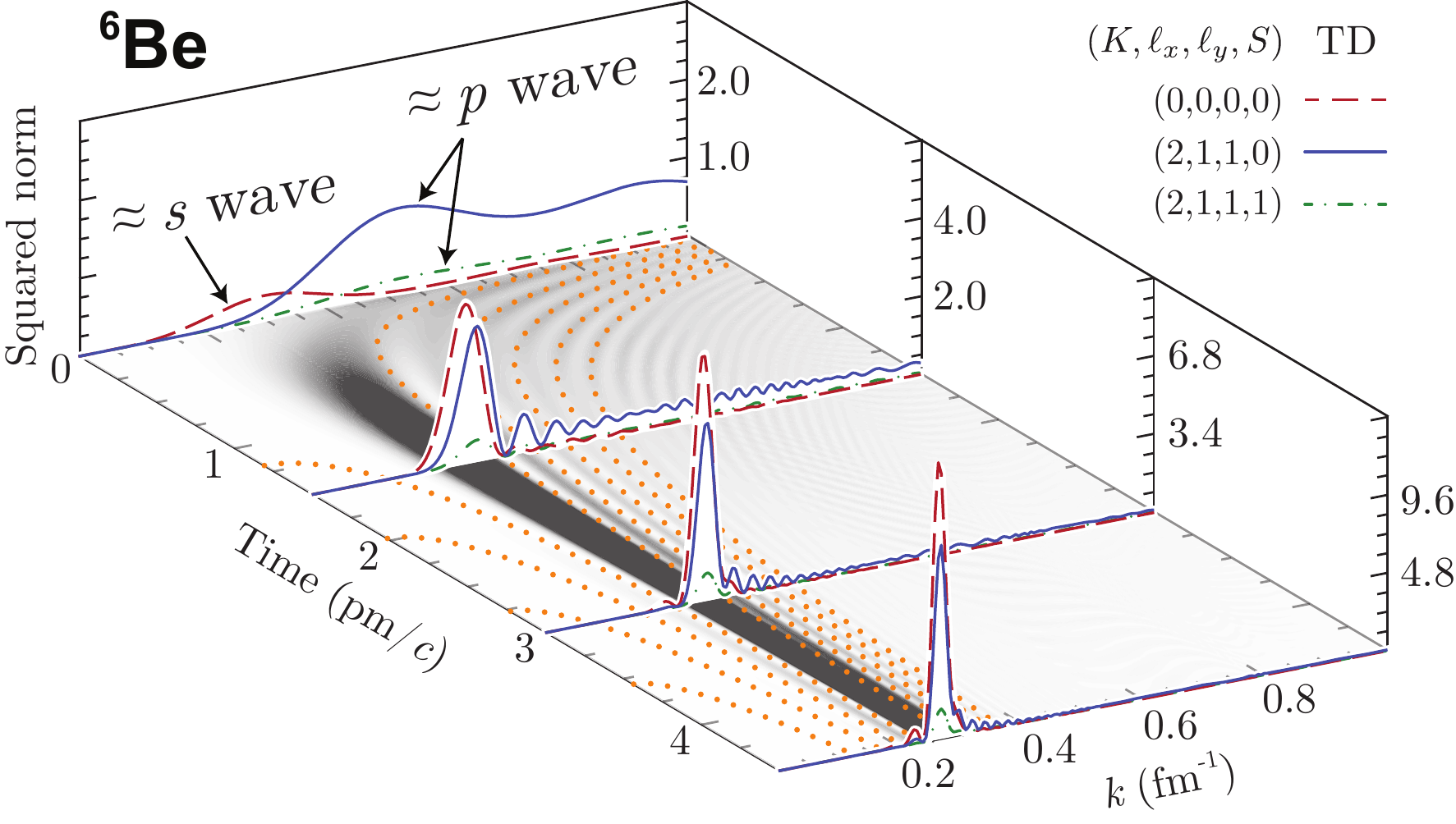}
\caption{Time evolution of the wave functions of ${^6}$Be. Configurations are labeled as $(K,\ell_x,\ell_y,S)$ in Jacobi-T coordinates. $K$ is the hyper-spherical quantum number, $\ell$ is the orbital angular momentum in Jacobi coordinate, and $S$ is the total spin of valence protons. The projected contour map represents the sum of all configurations in momentum space; the interference frequencies are denoted by dotted lines corresponding to different ${\mathfrak n}$-values. See Ref.\,\cite{Wang2021} for details.}\label{Wang2021_1}
\end{figure*}

An approximate treatment of $2p$ emission was proposed in Ref.\,\cite{Bertulani2008}, in which the c.m. motion for the two valence protons was described classically. In a more realistic case, the early stage of the $2p$ emission from the ground state of $^{6}$Be was well demonstrated using a time-dependent method in Refs.\,\cite{Oishi2014,Oishi2017}. Initially, ($t=0$) when the wave function is relatively localized inside the nucleus, the density distribution shows two maxima for $^6$Be; these are associated with the diproton/cigarlike configuration characterized by small/large relative distances between valence protons (see Fig.\,\ref{Wang2019}). Meanwhile, the different time snapshots of the density distribution $\rho$ and the corresponding density changes $\rho_{d}$ are shown in Fig.\,\ref{Oishi2017}. $\rho$ and $\rho_{d}$ are defined as
\begin{equation}
\begin{aligned}
&\rho=8 \pi^{2} r_{1}^{2} r_{2}^{2} \sin \theta_{12}\left|\Psi\left(t\right)\right|^{2}; \\
&\rho_d =8 \pi^{2} r_{1}^{2} r_{2}^{2} \sin \theta_{12}\left|\Psi_d\left(t\right)\right|^{2}/\langle\Psi_d(t) | \Psi_d(t) \rangle;\\
|&\Psi_{d}(t)\rangle = |\Psi(t)\rangle-\langle\Psi(0) | \Psi(t) \rangle|\Psi(0)\rangle.
\end{aligned}
\end{equation}
During the early stage of decay, two strong branches are emitted from the inner nucleus, as shown in Fig.~\ref{Oishi2017}. The primary branch corresponds to the protons emitted at small opening angles; this indicates that a diproton structure is present during the tunneling phase. This can be understood in terms of the nucleonic pairing, which favors low angular momentum amplitudes and therefore lowers the centrifugal barrier and increases the $2p$ tunneling probability  \cite{Grigorenko2009_2,Oishi2014,Oishi2017,Wang2019}. The secondary branch corresponds to protons emitted in opposite directions. This accords with the flux current analysis shown in Fig.\,\ref{Wang2019}.

To capture the asymptotic dynamics and improve our understanding of the role of final-state interactions, the wave function of $^6$Be has also been propagated to large distances (over 500\,fm) and long times (up to 30\,pm/$c$); this facilitates analysis of asymptotic observables including nucleon--nucleon correlation \cite{Wang2021}. After tunneling through the Coulomb barrier, the two emitted protons tend to gradually separate under Coulomb repulsion \cite{Oishi2017,Wang2021}. Eventually, the $2p$ density becomes spatially diffused, which is consistent with the broad angular distribution measured in Ref.\,\cite{Egorova2012}. This corresponds to an opposite trend for the momentum distribution of the wave function, in which the emitted nucleons move with a well-defined total momentum, as indicated by a narrow resonance peak at long times \cite{IdBetan2018}. Figure\,\ref{Wang2021_1} illustrates the dramatic changes in the wave function and configurations of $^6$Be during the decay process. The gradual transition from the broad to narrow momentum distribution exhibits a pronounced interference pattern, which is universal for two-nucleon decays and is governed by Fermi's golden rule. The interference frequencies, as indicated by dotted lines in Fig.\,\ref{Wang2021_1}, can be approximated by $(\frac{\hbar^2}{2m} k^2 - Q_{2p}) t = {\mathfrak n}\pi \hbar$, where ${\mathfrak n}$ = 1, 3, 5 $\cdots$ (i.e., they explicitly depend on the $Q_{2p}$ energy).

Moreover, the configuration evolution also reveals a unique feature of three-body decay. As seen in Fig.\,\ref{Wang2021_1}, the initial ground state of $^{6}$Be is dominated by the $p$-wave ($\ell = 1$) components, and the small $s$-wave ($\ell = 0$) component originates from the non-resonant continuum. As the system evolves, the weight of the $s$-wave component---approximately corresponding to the Jacobi-T coordinate configuration $(K,\ell_x,\ell_y,S)$ = (0,0,0,0)---gradually increases and eventually dominates, because it experiences no centrifugal barrier. Such transitions can also be revealed by comparing the internal and external configurations obtained via time-independent calculations \cite{Grigorenko2003}. This behavior can never occur in the single-nucleon decay, owing to the conservation of orbital angular momentum; however, it occurs in the two-nucleon decay because correlated di-nucleons involve components with different $\ell$-values \cite{Catara1984,Pillet2007,Hagino2014,Fossez2017}. In addition, for $2p$ decays, the Coulomb potential and kinetic energy do not commute  in the asymptotic region \cite{Grigorenko2009}, leading to additional configuration mixing.

\begin{figure}
\includegraphics[width=0.8\columnwidth]{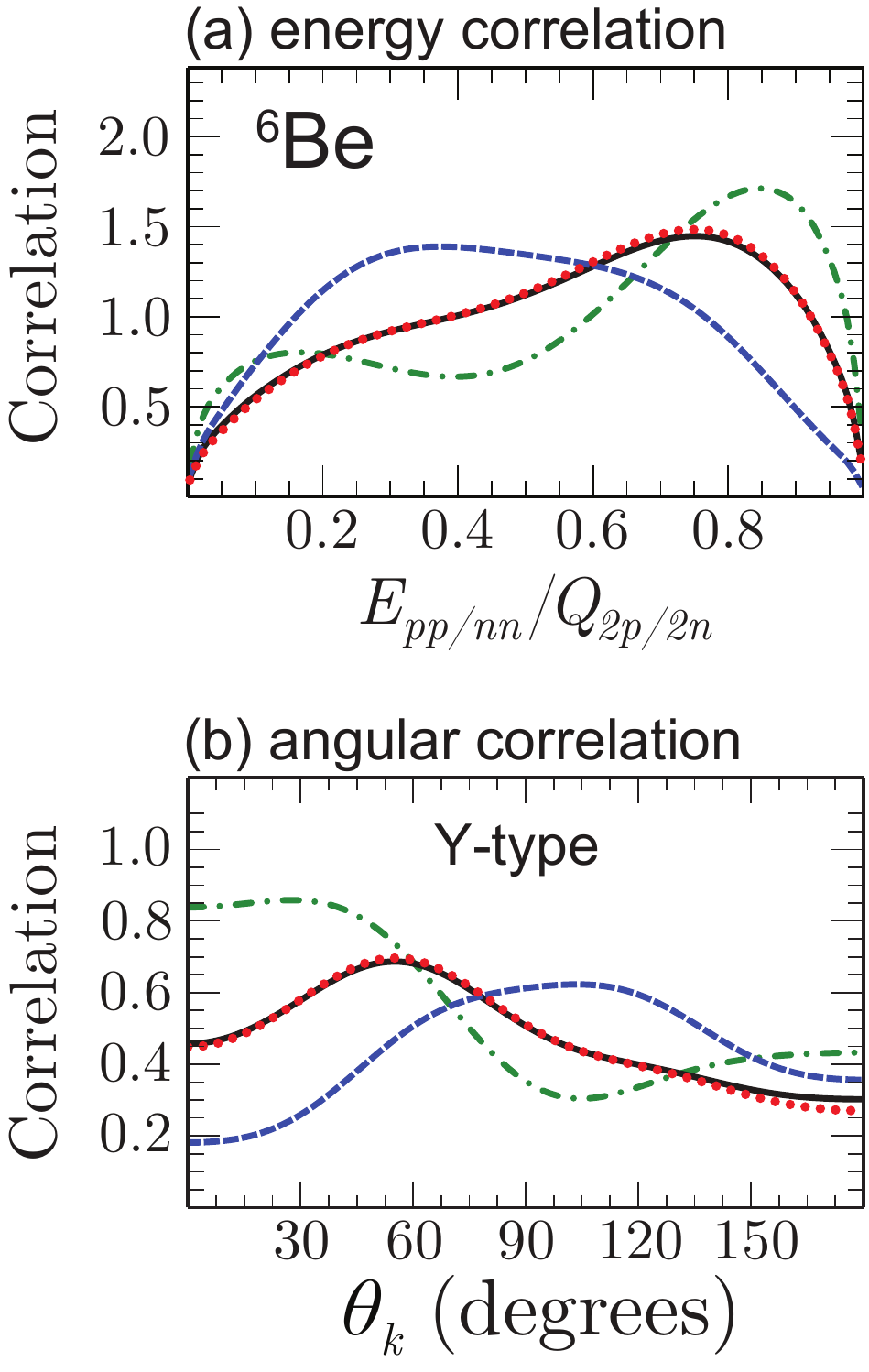}
\caption{
(a) Asymptotic energy and (b) angular correlations of nucleons emitted from the ground state of $^6$Be (top) and $^6$He$^\prime$ (bottom), as calculated at ${t=15}$\,pm/${c}$ with different strengths of the Minnesota interaction \cite{Thompson1977}: 
standard (solid line), strong (increased by 50\%; dashed line), and weak (decreased by 50\%; dash-dotted line). Also shown are the benchmarking results obtained by the Green's function method (GF; dotted line) using the standard interaction strength.
$\theta_k$ is the opening angle between $\vec{k}_x$ and $\vec{k}_1$ in the Jacobi-Y coordinate system, and $E_{pp/nn}$ is the kinetic energy for the relative motions of the emitted nucleons. See Fig.\,\ref{Wang2021} and Ref.\,\cite{Wang2021} for definitions and details.
}\label{Wang2021_2}
\end{figure}

It has been also found that the $2p$ decay width is sensitive to the strength of the pairing interaction \cite{Barker2003,Oishi2017,Wang2021}, because the diproton structure promotes the tunneling process. Moreover, Figure.\,\ref{Wang2021_2} shows the asymptotic correlations of $^{6}$Be as a function of the nucleon--nucleon interaction strength $V_{pp/nn}$ \cite{Wang2021}. The obtained results have been also benchmarked with the Green's function method, because the time evolution operator can be written as the Fourier transform of $\hat{G} = (E- \hat{H} +i\eta)^{-1}$:
\begin{equation}\label{Green_function}
	e^{-i \frac{\hat{H}}{\hbar} t}=\frac{e^{\frac{\eta}{\hbar} t}}{2 \pi i} \mathcal{F}\left(\hat{G} , E \rightarrow \frac{t}{2 \pi \hbar}\right).
\end{equation}
As shown in Fig.\,\ref{Wang2021_2}, the attractive nuclear force is not only responsible for the presence of correlated di-nucleons in the initial state but also significantly influences the asymptotic energy correlations and angular correlations in the Jacobi-Y angle $\theta_k$. This indicates that, even though the initial-state correlations are largely lost in the final state, certain fingerprints of the di-nucleon structure can still manifest themselves in the asymptotic observables.

\section{Open Problems}

Because the exotic phenomenon of direct $2p$ decay was discovered not long ago, only a few nuclei have thus far been found to exhibit it, which precludes systematic studies. Meanwhile, the corresponding microscopic theories remain under development. Although impressive progress has been achieved in both experiment and theory, many open problems remain under debate. Here, we briefly introduce some of them that are crucial for a better understanding of the $2p$ decay process; this might help us gain a deeper insight into the properties of open quantum systems.

\subsection{Connecting structure with asymptotic correlation}

As shown above, the asymptotic nucleon--nucleon correlation and other decay properties are strongly determined by the nuclear structure. Therefore, deciphering these observables can help us to understand the nucleus interior and the properties of nuclear forces. In particular, the study of short-range correlations reveals the fundamental structure of nucleonic pairs at very high relative momenta  \cite{Hen2017,Zhang2016,Huang2020,Shen2022}. Regarding the $2p$ decay, this observed long-range correlation corresponds to a low-momentum phenomenon, which indicates the interplay between the diproton condensate inside the nucleus and the Cooper pairing during decay \cite{Dobaczewski2007,Hagino2005}.

Although these asymptotic observables can be evaluated using various kinds of theoretical models, the precise connection between nuclear structure (including  pairing and continuum effects) is still largely unknown. One reason for this is the complicated open quantum nature of this three-body system, the other is the very different scales. The asymptotic nucleon--nucleon correlation and other observables are distorted during the decay process. This situation is exacerbated by $2p$ decay in the presence of long-range Coulomb interactions.

\begin{figure}
\includegraphics[width=1\columnwidth]{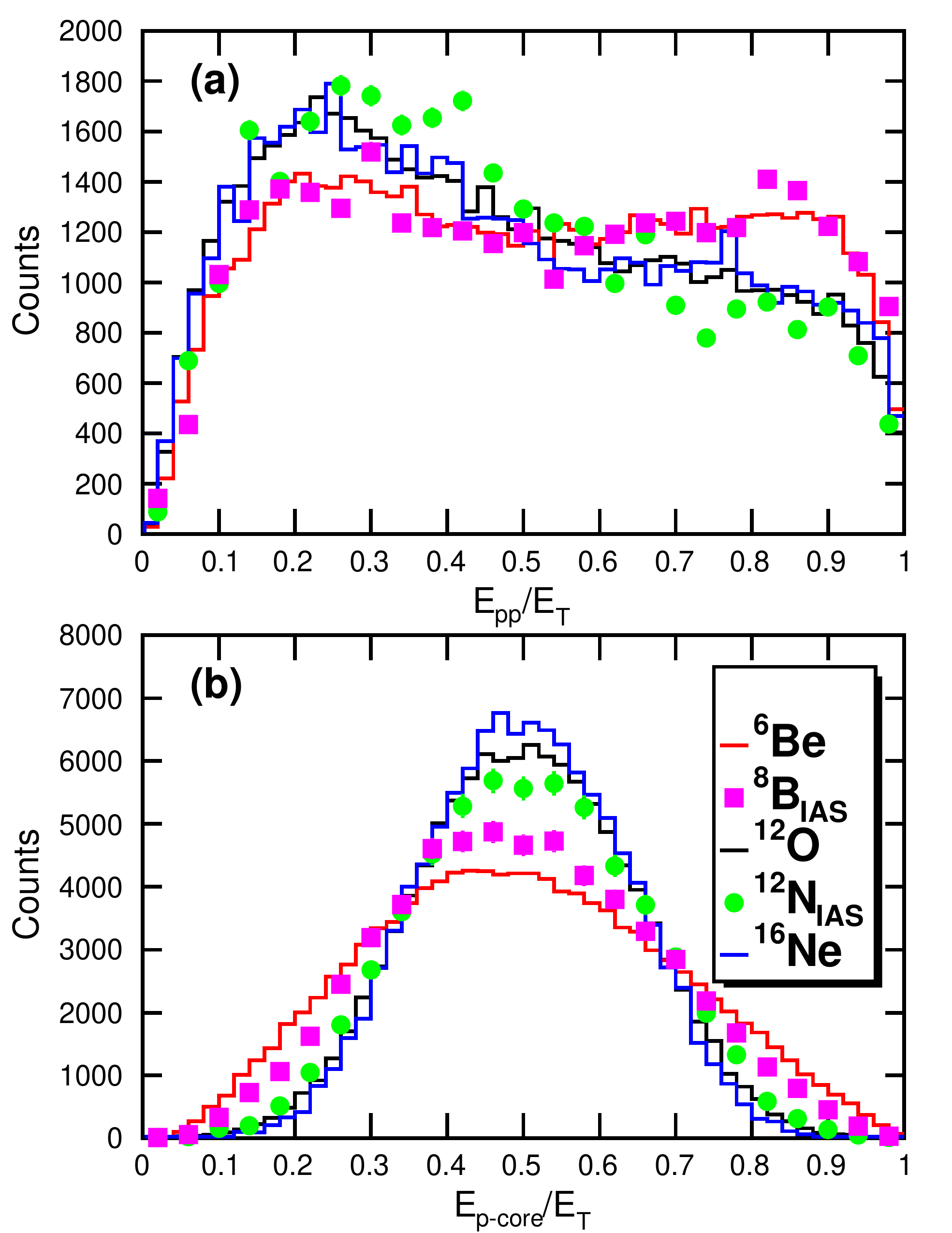}
\caption{Energy correlations of different $2p$ emitters in (a) Jacobi-T and (b) -Y coordinates. Ground-state $2p$ emitters are plotted as histograms whilst $2p$ emissions from isobaric analogs are plotted as data points. See Ref.\,\cite{Webb2019_2} for details.
}\label{Webb2019_2}
\end{figure}

In one pioneering work, $^{45}$Fe was---by analyzing the correlation of the emitted valence protons---found to contain a moderate amount of the $p^2$ component, \cite{Miernik2007,Pfutzner2012}. Later, the nucleon--nucleon correlations of $psd$-shell $2p$ emitters were systematically studied. As shown in Fig.\,\ref{Webb2019_2}, among these $2p$ emitters, the energy correlations of $^{12}$O and its isobaric analog $^{12}$N$_{\rm IAS}$ resemble those of $^{16}$Ne but differ notably from the energy correlations of $^{6}$Be and $^{8}$B$_{\rm IAS}$. This indicates that the valence protons of $^{12}$O might have more $sd$-shell components, even though, in the naive shell-model picture, these valence protons fully occupy the $p$-wave orbitals. Consequently, these qualitative analyses provide us with useful information about the valence protons and corresponding $2p$ emitters. Moreover, to gain deep insight into the connection and extract structural information from these asymptotic observables, high-quality experimental data and comprehensive theoretical models are required.

\subsection{Excited {\it 2p} decay}

As discussed, the phenomenon of $2p$ decay is not limited to the ground state: more and more excited nuclei have been found to exhibit $2p$ decay. Unlike ground-state $2p$ emitters, the $2p$ decay from excited states usually has a large decay energy $Q_{2p}$. This may lead to dense level density and more possible decay channels, which make it hard to determine the decay path and corresponding mechanism. Moreover, those high-lying states with large decay energies lie beyond the capabilities of many microscopic theories, in which the nuclear structure cannot be precisely described. So far, most tools in the market have dealt with this problem via assumptions and simulations. No self-consistent framework has yet been found that can be applied to comprehensively study these excited $2p$ emitters.

Meanwhile, nuclei can exhibit $\beta^+$-delayed $2p$ emission. For instance, one such decay process was observed from the ground state of $^{22}$Al \cite{WangYT2018}. The decay proceeds through the isobaric analogue state (IAS) of $^{22}$Mg to the first excited state in $^{20}$Ne, and it is confirmed by proton-gamma coincidences. This poses a big challenge for theoretical studies; this is because, even though protons and neutrons have tiny differences attributable to their inner structures and the Coulomb interaction, they are treated as identical particles in most theoretical models, by assuming isospin-symmetry conservation. However, this transition is considered to be forbidden by the isospins, because the $2p$ decay cannot change the isospin from  $T = 2$ in the case of $^{22}$Al and $^{22}$Mg$_{IAS}$ to $T = 0$ in the case of $^{20}$Ne. This indicates that isospin mixing is crucial in such $\beta^+$-delayed $2p$ emission processes; this requires further theoretical and experimental investigations.

\subsection{Multi-proton radioactivity}

\begin{figure}[!htb]
\includegraphics[width=1\columnwidth]{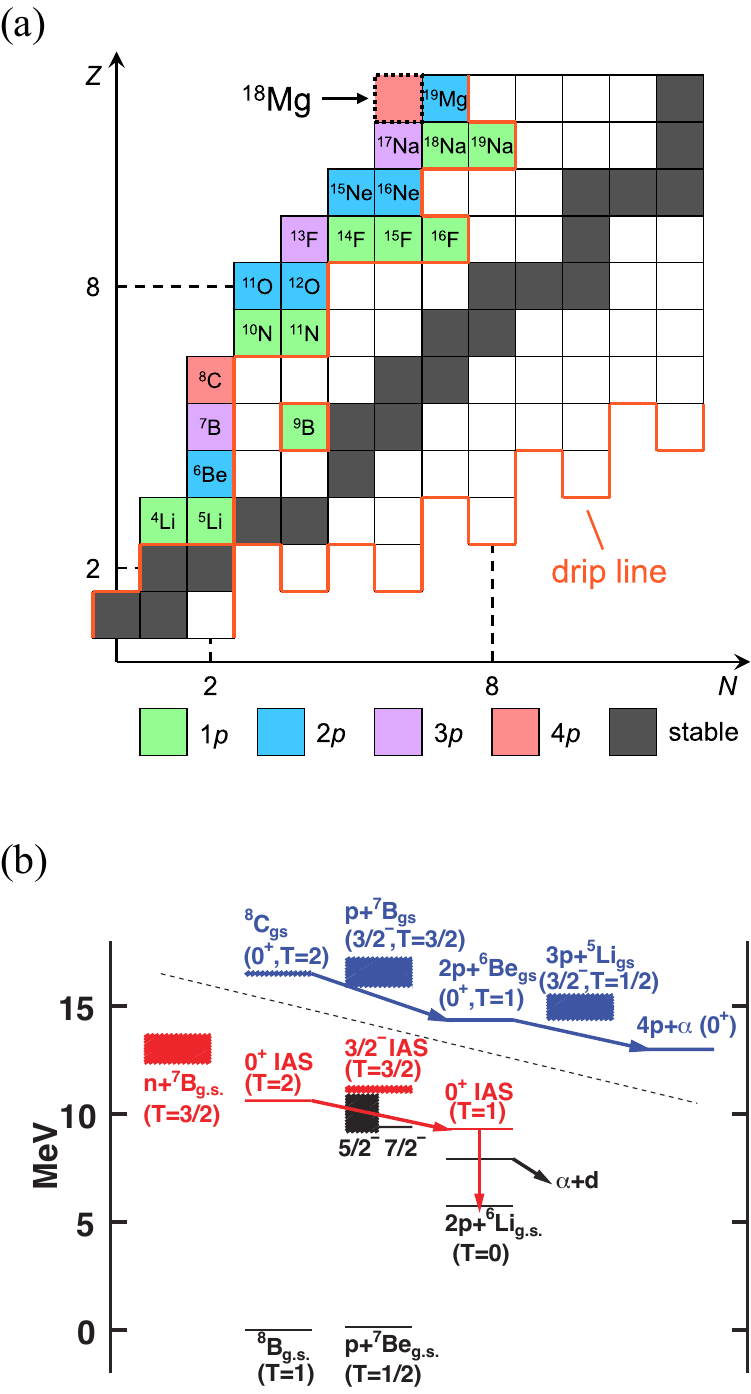}
\caption{(a) Subsection of the nuclear chart for multi-proton emission. These highlighted nuclei have been observed to decay via $1p$ (green), $2p$ (blue), $3p$ (purple), and $4p$ (pink) emissions. (b) The level diagrams for $^8$C and its isobaric analog $^8$B. The diagram for $^8$C is shifted up for display purposes. The full width at half maximum of the levels is indicated by the hatched regions. The isospin-permitted decays are presented in color. See Refs.\,\cite{Charity2010,Jin2021} for details.
}\label{Jin2021}
\end{figure}

Beyond the proton/neutron dripline, exotic decays may happen. $2p$ decay is just one of them. Recently, more and more exotic particle emissions (e.g., $3p$ or $4p$ decay) have been experimentally observed. The ground state of $^{31}$K has been found to exhibit a $3p$ decay \cite{Kostyleva2019}. Moreover, the newly discovered elements $^8$C and $^{18}$Mg are considered to exhibit $4p$ emissions \cite{Charity2010,Jin2021}. Owing to the large imbalance of the proton-neutron ratio, these $3p$ or $4p$ emitters are strongly coupled to the continuum and are more unstable compared to the $2p$ emitters.
The decay mechanism of these multi-particle emissions is an interesting phenomena to investigate. As shown in Fig.\,\cite{Jin2021}(b), $^8$C and $^{18}$Mg arguably feature a 2$p$+2$p$ decay mode, in which the intermediate states are the ground states of $^6$Be and $^{16}$Ne, respectively. However, owing to the low-statistic experimental data, their decay properties have not yet been fully determined. To address this problem, more physical observables are required. 

\begin{figure}[!htb]
\includegraphics[width=1\columnwidth]{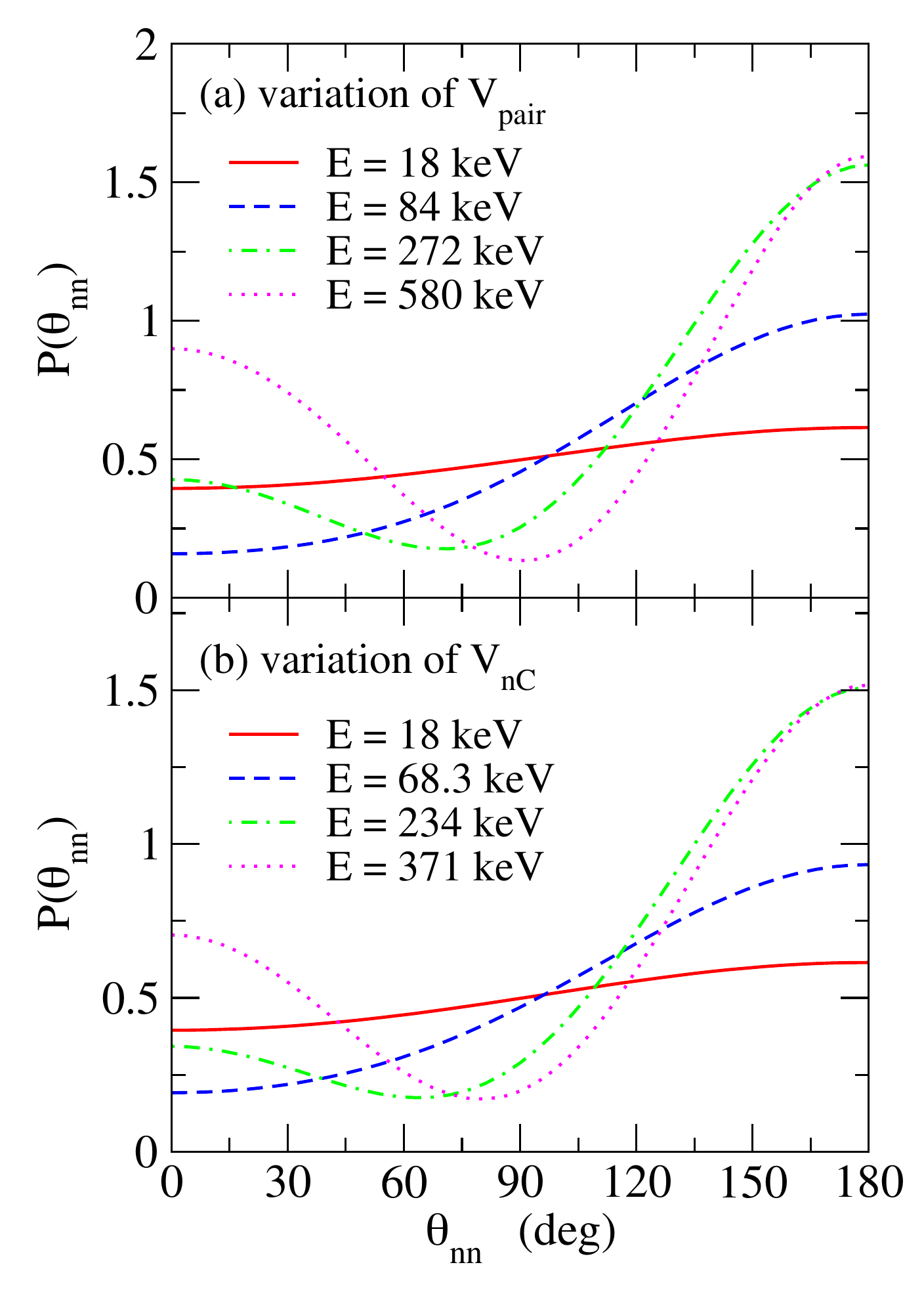}
\caption{The coordinate-space
angular density of the ground state of $^{26}$O as a functional two-neutron decay energy $E$. 
The upper panel is obtained by varying the energy (by changing the pairing interaction of 
the two valence neutrons). The lower panel is obtained by shifting the resonance energy of the $d_{3/2}$ state in $^{25}$O, keeping the strength of the pairing interaction identical. See Ref.\,\cite{Hagino2016} for details.
}\label{Hagino2016}
\end{figure}

\subsection{Multi-neutron emission and mirror symmetry breaking}

As the fermionic building blocks of a nucleus, the positively-charged proton and neutral neutron are almost identical in all respects except for their electric charge. This is a consequence of the isospin symmetry
\cite{Wilkinson1970,Warner2006}, which is weakly broken in atomic nuclei, primarily by the Coulomb interaction. The interplay between the short-ranged nuclear force and long-range electromagnetic force can be studied by investigating $2p$ and two-neutron radioactivity  \cite{Goldansky1960,Blank2008,Pfutzner2012,Wang2021}. However, the neutron dripline is harder to reach and detecting neutrons is more challenging than detecting charged particles; hence, the experimental information regarding two-neutron ($2n$) decay is limited.

So far, the weakly neutron-unbound $^{16}$Be and $^{26}$O are arguably the best current candidates for observing the phenomenon of $2n$ radioactivity \cite{Kohley2015,Kondo2016,Grigorenko2013,Hagino2014,Adahchour2017,Fossez2017,Casal2018,Grigorenko2018,Li2021}. Beryllium isotopes usually suffer large deformations, which might influence the $2n$ decay process of $^{16}$Be. As for $^{26}$O, its ground state is very near to the threshold, which might result in some unique phenomena.

Inside the nucleus, the initial $2n$ density of $^{26}$O also contains a dineutron, cigarlike, and triangular structure \cite{Hagino2014,Wang2017}. These three configurations are characteristic of the $d$-wave component. The very small $Q_{2n}=18\pm 5$\,keV value \cite{Kondo2016} in $^{26}$O  makes this nucleus a candidate for $2n$ radioactivity.  When approaching the threshold, the presence of the centrifugal barrier is expected to lead to changes in the asymptotic correlations. As shown in Fig.\,\ref{Hagino2016}, the angular correlation of $^{26}$O becomes almost uniformly distributed. This is because, for small values of  $Q_{2n}$, the $2n$ decay is dominated by the $s$-wave component, and the angular distribution becomes essentially isotropic \cite{Grigorenko2003,Grigorenko2018,Wang2021_2}. This asymptotic behavior might represent an interesting avenue of research for further experimental studies. 

When extending further beyond the neutron dripline, $^{28}$O is predicted to be a $4n$ emitter \cite{Grigorenko2011,Li2021,Brown2022}. It has been found recently that the four neutrons can exist transiently without any other matter \cite{Duer2022}. Although it remains under debate whether the observed peak is a resonance \cite{Sobotka2022,Kisamori2016,Pieper2003,Fossez2017_2,Deltuva2018,Li2019,Higgins2020}, it does show several correlations in the presence of nuclear media, which constitutes an interesting feature to be further studied in $^{28}$O and similarly proton-rich systems.

\section{Summary}

As one of the most recently discovered exotic decay modes, $2p$ radioactivity has attracted considerable theoretical and experimental attention. The massive processes have been desgined to help us gain insight into the $2p$ decay mechanism and its corresponding asymptotic observables. In this review, we briefly introduced a small fraction of the recent studies. In terms of experiments, by using the decay dynamics and lifetime, different methods and detectors (e.g., including in-flight decay techniques, implantation decay, and time projection chambers) have been developed. Moreover, new techniques such as waveform sampling are under development as an extension of the traditional methods. Meanwhile, the theoretical developments (starting from the configuration interaction and three-body model) are aiming to construct a unified framework in which the interplay of structural information and decay properties can be more fully understood.

With the establishment of next-generation rare isotope beam facilities, more and more $2p$ emitters (as well as other exotic decays) will be discovered. Along with the development of new detector techniques and self-consistent theoretical frameworks, the exotic mechanism, decay properties, and structural information regarding $2p$ decay will be comprehensively investigated, leading to a better understanding of nuclear open quantum systems.

\section*{Acknowledgements}

This work is partially supported by the National Natural Science Foundation of China (Nos. 12147101, 11925502, 11935001, 11961141003, 11890714), the National Key R\&D Program of China (No. 2018YFA0404404), the Strategic Priority Research Program of Chinese Academy of Sciences (No. XDB34030000), and the Shanghai Development Foundation for Science and Technology (No. 19ZR1403100).

%


\end{document}